%% file: templateArxiv.tex
\documentclass{article}

\usepackage{PRIMEarxiv}
\usepackage{amsmath}
\usepackage[utf8]{inputenc} 
\usepackage[T1]{fontenc}    
\usepackage{hyperref}       
\usepackage{url}            
\usepackage{booktabs}       
\usepackage{amsfonts}       
\usepackage{nicefrac}       
\usepackage{microtype}      
\usepackage{lipsum}
\usepackage{fancyhdr}       
\usepackage{graphicx}       
\graphicspath{{media/}}     
\usepackage{subcaption}
\usepackage{tikz}
\def\checkmark{\tikz\fill[scale=0.4](0,.25) -- (.20,0) -- (1,.5) -- (.20,.10) -- cycle;} 

\usepackage{multirow}
\usepackage{graphicx}
\usepackage{makecell}
\usepackage{enumitem}
\usepackage{algorithm}
\usepackage{algpseudocode}
\title{Predicting Group Choices from Group Profiles

}

\author{
  Hanif Emamgholizadeh\\
  Free University of Bozen-Bolzano \\
  Bolzano\\
  \texttt{hemamgholizadeh@unibz.it} \\
   \And
  Amra Deli\'{c} \\
  University of Sarajevo \\
  Sarajevo\\
  \texttt{adelic@etf.unsa.ba} \\
  \AND
  Francesco Ricci\\
  Free University of Bozen-Bolzano \\
  Bolzano\\
  \texttt{fmr959@gmail.com} \\
}

\begin{document}
\maketitle

\begin{abstract}
Group recommender systems (GRSs) identify items to recommend to a group of people by aggregating group members' individual preferences into a group profile, and selecting the items that have the largest score in the group profile. The GRS predicts that these recommendations would be chosen by the group, by assuming that the group is applying the same preference aggregation strategy as the one adopted by the GRS. However, predicting the choice of a group is more complex since the GRS is not aware of the exact preference aggregation strategy that is going to be used by the group. 

To this end, the aim of this paper is to validate the research hypothesis that, by using a machine learning approach and a data set of observed group choices, it is possible to predict a group's final choice, better than by using a standard preference aggregation strategy. Inspired by the  Decision Scheme theory, which first tried to address the group choice prediction problem, we search for a group profile definition that, in conjunction with a machine learning model, can be used to accurately predict a group choice. Moreover, to cope with the data scarcity problem, we propose two data augmentation methods, which add synthetic group profiles to the training data, and we hypothesize they can further improve the choice prediction accuracy. 

We validate our research hypotheses by using a data set containing 282 participants organized in 79 groups. The experiments indicate that the proposed method outperforms baseline aggregation strategies when used for group choice prediction. The method we propose is robust with the presence of missing preference data and achieves a performance superior to what humans can achieve on the group choice prediction task. Finally, the proposed data augmentation method can also improve the prediction accuracy. Our approach can be exploited in novel GRSs to identify the items that the group is likely to choose and to help groups to make even better and fairer choices.
\end{abstract}


\input{01_Introduction}
\input{02_SOTA}

\input{03_Theory}
\input{04_Model}
\input{05_Experiments}

\input{06_Results}

\input{07_Discussion}
\bibliographystyle{unsrt}  
\bibliography{references}

\end{document}

%% file: 01_Introduction.tex
\section{Introduction}
\label{Sec:Intro}

Recommender Systems (RSs) are information retrieval tools that help their users to make better decisions by suggesting items that are likely to meet their needs and wants~\cite{ricci:2022:ch1}. Group Recommender Systems (GRSs) are special types of RSs aiming at identifying items that, if experienced by a group of people, will satisfy, as much as possible, all the group members~\cite{MASTHOFF2022}. Group recommendations are constructed by algorithms that leverage preference aggregation strategies, which combine, for each item, the item preference scores of the group members into a single score, for instance, by averaging the group members' preference scores~\cite{JAMESON04}.   

While the ultimate goal of a GRS is to generate useful recommendations for a group, the system may benefit from a component that, by relying on the knowledge of individual preferences, generates a prediction of the more likely group's choice. That prediction may be used directly as a recommendation, helping the group to quickly converge to that decision. However, the group choice prediction can also be exploited for generating alternative and better choices. For example, the GRS could identify options that are similar to the predicted choice but fairer, which can be achieved, for instance, by reducing the variance of the group members' predicted satisfaction scores.

In this work, we focus explicitly on the problem of group choice prediction, and we propose a machine learning-based solution that leverages a training data set of observed groups. The groups are described by their group profiles, which are constructed with a preference aggregation strategy. We then aim at predicting the choices of groups not present in the training set. We note that while the group choice is typically the result of the group decision-making process, we aim at predicting it solely from the knowledge of the group members' preference data, which are aggregated in the group profile. 
Our approach is inspired by the Social Decision Scheme (SDS) theory~\cite{stasser1999}. SDS theory builds a group profile, named {\it group preferences composition}, only on the base of the individual preferences (the group members' preferred options), and it assumes that a so called \textit{Social Combination Process} or \textit{Social Decision Scheme} summarizes the relationship between the initial group preferences' composition and the final group choice, that is the collective response.

We note that classical preference aggregation techniques, which are used in GRSs, can generate group choice prediction: the item with the largest aggregated preference score is predicted as the choice of the group. Hence, they also follow SDS assumption that the group choice solely relies on the individual preferences of the group members. However, they  {\it mechanically} dictate the group choice by using hard-coded rules, such as: ``the item with the largest average individual rating must be the group choice''. Moreover, which preference aggregation strategy must be used for predicting a specific group's choice is generally unknown, since real groups may use a variety of different preference aggregation strategies~\cite{MASTHOFF04,DELIC2018GDM,FORSYTH14}. Conversely, SDS theory suggests to reconstruct the social decision scheme and the group choice by using observational data describing how, in groups, the members' preferences determine the corresponding groups' choices. 

In that respect, by having at our disposal a data set containing information about group members' individual preferences and the corresponding post-interaction group choice, and by extending the SDS theory, we combine the heuristics of an aggregation strategy with a machine learning model (multinomial logistic regression) to predict the effect of the inter-group interaction. In other words, from a data set of real groups containing the knowledge of the group members' preferences and final group choices, we build a prediction model of the group's choice from the group members' individual preferences aggregated in a group profile by a preference aggregation strategy. 

Moreover, in order to cope with the (typically) limited number of observed groups and their choices that are in the training set, we conjecture, by following standard machine learning approaches, that it is possible to improve the accuracy of the choice prediction model with two, specifically designed, data augmentation methods~\cite{wong2016}. The objective is to bring some additional knowledge about typical decision-making behaviors in groups by enriching the training set with synthetic but likely to be observed groups (profiles) and their corresponding choices. In the first method, we add synthetic group profiles, which are called \textit{Winners}, and represent cases where all the group members prefer an option and the group (consequently) chooses that option. The second type of synthetic group is called \textit{Permutations}. These groups have profiles (and choices) that are obtained from the profiles of real, observed groups by making a permutation of the options and accordingly their scores; for instance, the group score of the first and second options in a real, observed group are swapped in a permutation group. If the group choice, as assumed by SDS, depends only on the scores of the options in the group profile, the choice of the group with the permuted profile will be the option obtained by the permutation of the originally chosen option. Hence, in the example mentioned above, if in the original group, the choice was the first option, in the permuted group the choice must be the second option. 

We have tested the effectiveness of our learning approach for group choice prediction on a data set describing the preferences and the choices of 282 participants organized in 79 real groups while deciding on which travel destination to visit together.  The precise formulation of our research hypotheses is in Section~\ref{sec:hyp}, and hereby we summarize the main results. We show that the proposed learning approach, which is named Learning-based Choice Prediction (LCP), generates significantly better predictions of the groups' choices in comparison to those based on classical preference aggregation strategies (called PACP - Preference Aggregation-based Choice Prediction). That result holds even when only partial information of the group members' preferences is available, i.e., when the system misses some group members' ratings. Our method has also a better prediction accuracy compared to what is achieved by humans, when, after having observed the group members' preferences, they predict the likely choice of the group. Moreover, we show that by using data augmentation (winners and permutations), the prediction performance can be improved and the predicted distribution of group choices can be made more similar to the observed distribution of the group choices. 
In summary, the main contributions of this paper are: 
\begin{itemize}
    \item A novel learning approach (LCP) to predict group choices given the knowledge of the group members' preference that significantly outperforms the accuracy of baseline preference aggregation strategies (PACP), that is robust with respect to missing preference data and that also outperforms human-based group's choice prediction.
    
    \item A data augmentation method that, by adding synthetic group profiles (winners and permutations), improves the group choice prediction accuracy of LCP and makes the distribution of the predicted group choices more similar to the observed one (ground truth).
    
\end{itemize}

We stress the practical importance and value of the obtained results. The prediction of a target group choice can be used by a GRS to indicate which option is the current inclination of the group, hence helping the group to quickly come to such a decision. Moreover, by having the knowledge of the likely choice of the group, the GRS can leverage this information to generate other recommendations, for instance, presenting items similar to the predicted choice but with additional important properties, for instance, items that are more novel or fairer choices. Hence, we believe that our results can open the research on novel and effective group recommendation techniques, and especially conversational ones, which can greatly benefit from the prediction of the likely choice of a group to better interact with the group members in supporting their decision-making process.

The rest of the paper is structured as follows. In the next Section, we provide an overview of the related work on group recommender systems. In Section~\ref{sec:BT}, we compare the proposed approach with the state of the art and we formulate our research hypotheses. In Section~\ref{sec:model}, the group profile generation mechanism and the choice prediction learning approach are elaborated in details. In Section \ref{sec:data_augm}, data augmentation is discussed and our approach based on the generation of synthetic group profiles is presented. In Section~\ref{sec:experiments}, the evaluation procedure is explained, and then in Section~\ref{sec:results}, the results supporting our two research hypotheses are presented. Finally, in Section~\ref{sec:conclusion}, we summarise the paper contribution, discuss limitations of our approach, and indicate lines of future work.

%% file: 02_SOTA.tex
\section{State of the Art}

\label{sec:sota}

In this section, we first survey approaches that deal with the construction of the group profile which have been developed in the Group Recommender Systems (GRSs) literature (Subsections~\ref{sec:grs_profiles}, \ref{sec:aggr_strat}, and~\ref{sec:ml}). Then, in Subsection~\ref{sec:sds}, we present an approach to group profile generation, namely Social Decision Scheme, specifically introduced for predicting the group choice.

\subsection{GRS based on group profile}
\label{sec:grs_profiles}

Group Recommender Systems (GRSs) are designed to find items whose joint experience in a target group would be satisfactory for all the group members~\cite{MASTHOFF2022,felfernig2018group}. Group recommendation methods can be divided into two main classes: combining recommendations and combining user profiles. In the first class, recommendations are first generated for each group member, and then based on them, group recommendations are selected. In the second class of methods, a group profile is first constructed by using preference aggregation techniques applied on the group members' individual profiles (preferences for the items, e.g., ratings). Then group recommendations are generated on the base of the constructed group profile. In this section, we focus on the second class of methods, profiles aggregation, as our choice prediction approach leverages this construct. 

\begin{figure}[!tbp]
    \centering
    \includegraphics[width=\linewidth]{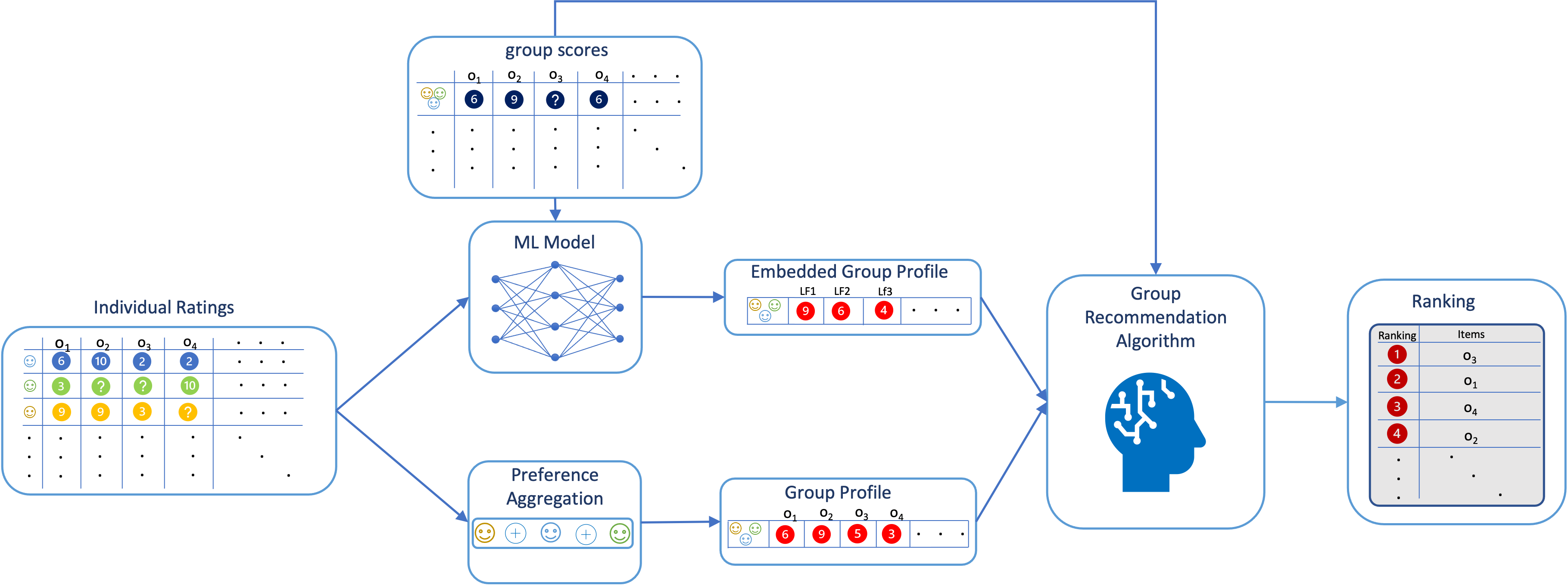}
    \caption{Schema of GRSs that utilize either standard preference aggregation strategies (lower workflow) or ML models (upper workflow). In the preference aggregation-based approaches, the system receives the \textit{individual rating} (actual or predicted) as input to construct \textit{group profiles}, by using a \textit{preference aggregation} strategy. The constructed group profiles contain predicted group scores for the options ($o_1, o_2, o_3,$ and $o_4$). This \textit{group profile} is used by the group recommendation algorithm to generate recommendations. In ML-based approaches, the system leverages \textit{individual ratings} and \textit{group scores} as input to construct the \textit{embedded group profile}. Embedded profiles are defined by latent features (e.g., LF1, LF2, and LF3). The \textit{embedded group profile}, in addition to \textit{group scores}, is used by the \textit{group recommendation algorithm} to generate recommendations.}
    \label{fig:GRSS}
\end{figure}

Figure \ref{fig:GRSS} shows the general schema used by GRSs to build a group profile and then group recommendations. There are two main paths to create a group profile: by using a preference aggregation strategy and by using Machine Learning (ML) techniques. The approaches based on the preference aggregation strategies take as input the individual preferences of the group members (\textit{individual ratings}) and construct for each group a profile by relying on a selected \textit{preference aggregation} strategy. Then a recommendation algorithm leverages the target group profile, together with other group profiles, to generate recommendations. Conversely, ML methods create a group profile by taking as input individual ratings but also the available group choices or group scores (also called ratings by some authors), which indicate to what extent a set of observed groups like some evaluated options. Then, an ML model produces an \textit{Embedded Group Profile}, which is defined by hidden features of the group. In fact, generating group profiles is not the main goal of the ML models; this is a byproduct of the method used to predict the group scores. Hence, while the group profile generated by preference aggregation strategies contains the estimated group scores for the options, the group profile constructed by ML models contains latent features, and the value of each entry indicates the estimated importance of that dimension for the group representation. In both approaches, a \textit{group recommendation algorithm} uses the generated group profiles as input to compute recommendations, which are obtained by exploiting a range of different solutions, such as collaborative filtering or neural networks.

In the following sections, we will detail the two approaches to group profile generation, i.e., the one based on preference aggregation and the one leveraging ML.

\subsection{Preference Aggregation Strategies}
\label{sec:aggr_strat}

Group modeling or creating group profiles is likely the most essential aspect of aggregation-based Group Recommender Systems (GRSs). A large part of the research on GRSs has been dedicated to understanding how group members' individual preferences {\it are} (descriptive) or {\it should be}   (normative) aggregated to create group profiles. In~\cite{MASTHOFF04} a number of preference aggregation strategies, motivated by Social Choice theory, are described.
\begin{itemize}
\item \textit{Additive} - group members' individual ratings for each item (if available) are summed up to create a vector of group scores, one score for each item. Possible implementations of the additive strategy calculate the mean value (i.e., average strategy) or the median value (i.e., median strategy) of the individuals' ratings. 
\item \textit{Borda count} - each group member creates a ranked list of options according to his/her preferences; points are assigned to options, separately for each individual, based on the position of an option in a list (i.e., the last option gets zero points, the second last receives one point, etc.); a group score for an option is calculated as the sum of the individually assigned points.

\item \textit{Multiplicative} - individuals' ratings are multiplied to create group scores. 

\item \textit{Least misery} - an item's group score is the minimum of the individuals' ratings for the item; the strategy assumes that a group is as satisfied as its least satisfied member. 

\item \textit{Copeland Rule} - An item's group score is equal to the number of times that option beats other options minus the number of times it loses with respect to the other options. Option $i_1$ beats option $i_2$ in a group if the number of group members who prefer $i_1$ to $i_2$ is larger than the number of members who prefer $i_2$ to $i_1$.

\item \textit{Majority}: group score for each option is equal to the number of group members that have chosen the option as their individual or group choice.
\end{itemize}

Additive-based aggregation strategies typically treat all group members as equally important. However, there are situations where certain group members may have different levels of importance. Weighted-sum or weighted average aggregation strategies can be used to address this problem. For instance, Ardissono et al. \cite{ardissono2003intrigue} used the weighted sum aggregation strategy to construct a group profile that reflects the importance of individual group members. In this approach, each member is assigned a different weight to reflect their importance in constructing the group profile. Another type of aggregation strategy is the \textit{Distance-based} strategy~\cite{zhiwen2005adaptive}, which aims to minimize the total distance between the constructed group profile and the individual profiles of group members. In other words, given individual preferences (for instance individual ratings), the distance-based aggregation strategies create the group profile that minimizes the total distance of the constructed group profile from the individual group member's profile.

Another research direction for constructing group profiles uses individual rankings, instead of ratings. Similar to the distance-based strategies mentioned earlier, these methods aim to create a group profile that minimizes the constructed group profile's (ranking) distance from individual preferences (rankings). For instance, in~\cite{cook1978priority}, the authors propose a method for constructing group profiles that minimize the total distance to the profile (ranking) of individual members. Additionally, in~\cite{dong2021}, it is argued that group members may sometimes categorize potential options into approved (items they would like to consume) and disapproved (items they would not consume) groups individually. They introduced the concept of a preference-approval structure, which combines ranking and approval data to incorporate approved and disapproved alternatives during preference modeling. The authors proposed a group preference aggregation model that minimizes the total distance to individual preference-approval structures.

\subsection{Machine Learning Models}
\label{sec:ml}

Standard preference aggregation strategies, and their extensions, construct group profiles in a pre-defined and mechanical way. As a consequence, the constructed group profiles may not be optimal in different contexts. For instance, different groups might employ different strategies in order to reach satisfying decisions. Moreover, even when the same group is deciding on different options, they might employ different techniques to consider and evaluate these options. To overcome this problem, ML-based variants propose more adaptive models.

In \cite{cao2018AGREE} the authors propose a method that learns a group profile by using an attentive neural network based on existing individual user-item as well as group-item interactions \footnote{By an individual user-item interaction, the user's choice of that particular item is usually considered. Similarly, by a group-item interaction, a group choice of that item is usually considered.}. Hence, in this case, aside from individual user preferences, group preferences must also be available to the model. The constructed group profiles are used by another ML component for generating recommendations. The performance of this approach deteriorates when applied to ephemeral groups (created for one event only, or put in other words, groups for which there is only one group-item interaction entry) \cite{Sankar2020GroupIM}. To overcome this problem, the authors in~\cite{Sankar2020GroupIM} proposed an alternative solution. Here, the attention mechanism is extended with another neural approach that maximizes the Mutual Information between the group and its members. This assigns a greater weight to a user in the group profile when their current group is more similar to the ephemeral groups that the user belonged to in the past. We note that these methods follow the upper workflow indicated in Figure \ref{fig:GRSS}. They utilize known individual, as well as group preferences (stored in a dataset) for constructing group profiles and predicting group scores for items.

\subsection{Social Decision Scheme}
\label{sec:sds}

In the previous subsections, we have focused on the generation of a proper representation of the group preferences, called group profile, and the corresponding recommendation methods. However, as we mentioned in the introduction, when the goal is to support group decision-making it could be useful to predict the current inclination of a group for a specific option (group choice). 

The previously discussed preference aggregation strategies can be used to make predictions about the group choice: the option with the largest score after the strategy is applied is the choice that the group should make if the group adopts that strategy. However, it is worth mentioning that the main focus of these methods is not to predict what a group would choose given the individual preferences of the group members but to find ``the best'' option for the group, under certain constraints or goals that the aggregation mechanism aims for~\cite{sen1977social}. In other words, the methods indicate what the group should choose in order to fit certain constraints and goals.  

To the best of our knowledge, the Social Decision Scheme (SDS) theory, which is originally presented in~\cite{stasser1999} and also discussed in~\cite{friedkin_johnsen_2011}, is unique in explicitly focusing on the group choice prediction problem. SDS models how the collective response (group choice), which is the result of possibly complex inter-group interactions that happen during the group decision-making process, can be generated by relying only on individual preferences. Here, the group choice might not be the ``best'' option, according to a predefined mechanical approach, but the choice that the group reached according to their internal agreements. SDS proposes a set of basic modeling elements: (i) individual preferences, (ii) distinguishable distribution of group members' preferences, which corresponds to the group profile in the terminology used in this work as well as in GRSs, (iii) patterns of group influence (decision scheme), and (iv) collective responses (group choice). The theory states that individual preferences are the ingredients of the group profile (distinguishable distribution), and consensus processes act on such a group profile to yield a collective response (group choice)~\cite{stasser1999}. SDS claims that the social decision scheme (patterns of group influence), which describes the association between group profiles and the possible group responses (group choice), should be learned from data. Hence, in SDS theory, the group profile is generated from individual preferences, as in classical GRSs, but then an adaptive social decision scheme model, which could be learned from available data or is given a priori, is applied to yield the group response (group choice).

In Stasser's original approach, the entries of a group profile are constructed by counting the number of times an option has been indicated as the most preferred one by the group members - and such a group profile is called the distinguishable distribution (of the group preferences). Specifically, in his method, there are three essential elements. The first one is the set $P$ which contains all possible distinguishable distributions (group profiles). The number of possible distinguishable distributions, when $n$ is the number of available options and $r$ is the group size, is $_{(n+r-1)}C_r$, where $_mC_r$ is the binomial coefficient (the number of combinations, or the number of ways ways that $r$ elements can be drawn out from a set of $m$ elements). For instance, if a group of 2 members ($r = 2$) considers 2 options ($n=2$) one can observe $_3C_2 = 3$ distinguishable distributions: 1) the two members prefer the first option; 2) both members prefer the second option; 3) one member prefers one option and the other member the other option. The second element of the model is the $\pi$ vector, whose elements are the probabilities of observing each of the distinguishable distributions (group profiles), given another vector describing the apriori probabilities to observe the preference of individuals for the options. The last element of Stasser's model is the $D$ matrix, i.e., the social decision scheme matrix. The rows of this matrix correspond to the different group profiles and the columns correspond to the possible group responses (choices). Each entry $D_{ij}$ of the matrix is the probability that the $i$-th group profile leads to the $j$-th response. Stasser proposes that this matrix should be learned from data. A severe limitation of Stasser's model is that the number of distinguishable distributions (possible group profiles) grows rapidly with the number of options and fitting the matrix $D$ becomes practically impossible.

%% file: 03_Theory.tex
\section{Research Gap and Hypotheses}
\label{sec:BT}

\subsection{Discussion of the State of the Art}

\begin{figure}[!tbp]
    \centering
    \includegraphics[width=0.9\linewidth]{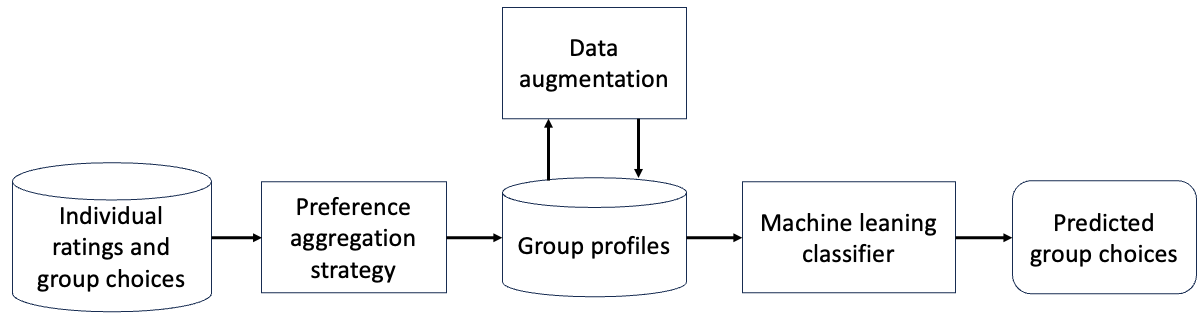}
    \caption{The logical schema of the proposed approach for learning the group choice from the group members' individual ratings of the considered options.}
    \label{fig:learningschema}
\end{figure}

Inspired by the SDS theory, we are tackling the group choice prediction task, i.e., to estimate what a group would actually choose from a limited set of options, by learning the choice function from a data set of observed group choices. The schema of our approach is shown in Figure~\ref{fig:learningschema}. Hence, the input of the proposed learning process is a set of groups, with the information of the members' individual ratings of the options, and the corresponding groups' choices. The result is a predictive model that, given a group profile, predicts the likely group choice. The proposed model can treat ephemeral groups, it does not use any possibly available information that an individual is part of more than one group, and any additional information about group members and groups, such as demographic or role data. 

In our work, we generalize the distinguishable distribution model of a group members' preferences that is introduced by SDS. We build a diverse set of potentially usable group profiles: by applying a range of preference aggregation strategies. The goal is to test whether the specific strategy used for aggregating the preferences of the group members may have an impact on the quality of the choice prediction. In our generalization of SDS theory, we also deal with the case that some individual preferences may not be available, hence we deal with missing data in the construction of group profiles. Moreover, we follow the original SDS idea that the association between group profile and group response (group choice) should be learned from data. However, we introduce a method to directly predict the group choice, without approximating the social decision scheme matrix $D$. The proposed learning method can be used with any vector-based representation of the group profile. We believe that our extension can also benefit other applications of SDS theory.

Classical preference aggregation strategies presented in Section~\ref{sec:aggr_strat}, as we previously noted, can be used for predicting a group choice as well. However, these strategies operate in a mechanical way and are not able to learn how the group choice depends on the computed group profile. Our method instead can achieve a better prediction accuracy because it learns the effect of inter-group interactions, as a map between the preferences, aggregated into a group profile, and the final group choice. For that reason, we apply machine learning to a range of diverse types of group profiles, which are computed by preference aggregation techniques, to understand what group profile may be better used to learn the map between profile and choice.

Similarly to the ML-based models that we have surveyed in Section~\ref{sec:ml}, our approach uses a training set of groups and choices, but, unlike them, we predict the group choice, not the group scores for the items. We believe that group scores are artificial concepts. In reality, a group makes a choice, and group members do not need to agree on their joint/common ratings of choices. We also note that the presented ML-based models use very sparse individual preferences of users for a huge set of items, and one of these approaches makes use of repeated evaluations of groups for items. Conversely, our method requires more dense individual preference data but it is able to work with ephemeral groups with only one observed choice for each group. 

\subsection{Research Hypotheses}
\label{sec:hyp}

In order to evaluate our proposed approach, we are interested in validating two hypotheses related to the task of predicting a group choice, over a limited set of options, after the group members have formulated their individual preferences and have interacted to make a joint decision to select one of the options (choice):

\begin{itemize}
    \item[\textbf{H1}] By using a proper machine learning model, trained on a data set of group choices, it is possible to effectively predict the choice of a group, by using only the individual preferences of the group members, encoded in a group profile.
\end{itemize}  

With this hypothesis, we aim to evaluate the general validity of the SDS theory, and our extension, in predicting a group choice. In fact, our objective is to examine whether it is possible to learn the social influence that happens during the group discussion utilizing solely individuals' preferences and groups' choices. To the best of our knowledge, this paper is the first attempt to validate the SDS theory using a real-world data set. 

To perform a quantitative assessment of this hypothesis we will compare the proposed approach to the choice prediction that preference aggregation strategies can generate. We hypothesize to be able to improve the prediction accuracy of these aggregation strategies. Moreover, the model proposed in~\cite{stasser1999} assumes the existence of all individual group members' ratings. However, in reality, in most cases, the user-option matrix is sparse, therefore, we also aim to examine the validity of this hypothesis when the user-option matrix is sparse. Finally, to make a further quantitative comparison of the achieved prediction quality we conduct a user experiment where humans are requested to predict the choices of the same groups, by only knowing the group members' preferences. We hypothesize that our method is competitive with the performance of human group choice prediction.

When trying to create a new model for recommending items to groups or predicting group choices, one faces the challenge of limited data availability. The majority of existing data sets are either small in size or lack clarity regarding their collection procedure, the assignment of \textit{group scores}, and the actual decision-making procedures followed by the groups. To this end, we hypothesize the following: 

\begin{itemize}
    \item[\textbf{H2}] In order to cope with the data scarcity of group choices, it is possible to use data augmentation, relying on synthetic group profiles and their choices, and further improve the quality of the proposed group choice prediction method. 
\end{itemize}

The precise data augmentation method that we propose is based on artificially introducing, only in the training set, groups along with their individual members' preferences, and group choices, that are not recorded in the data set. These synthetic groups, even if not actually observed, are likely to be ``observable''. For instance, even if in the data set there is not a group where all group members prefer one option to the other options, it can be assumed that if such a group exists in the data set, the individually preferred option will also be the group choice. 

In order to test the introduced hypotheses, we reuse a data set of group choices generated in a previously performed user study. Then, we conduct a set of extensive experiments. Both aspects are in detail explained in section \ref{sec:experiments}. In the following section, we provide a comprehensive description of the proposed ML-based choice prediction model and its comprising parts.

%% file: 04_Model.tex
\section{Group Choice Prediction Model}
\label{sec:model}

\subsection{Group Profiles}
\label{sec:genprofile}

We start with the definition of the important notations and the precise problem formulation. Let $U = \{u_1, \ldots, u_m\}$ be a set of users,  $O = \{o_1, \ldots, o_n\}$ a set of items or options, and $R = (r_{i,j})$ the $m \times n$ user-option rating matrix: $r_{i,j}$ is the rating (non-negative real number) given by user $u_i$ to option $o_j$. All notation utilized in the current and subsequent sections can be found in Table \ref{tab:notation}. To simplify the notation, we will refer to user $u_i$ as the $i$-th user, or even as user $i$. The same approach will be used for options. The rating matrix is in general partially defined, i.e., some of the ratings may be unknown. The user $u_i$'s profile $\boldsymbol{u}_i$,  is a real-valued vector formed by the $u_i$'s ratings:
\begin{equation}
    \label{Eq:user}
    \boldsymbol{u}_i = (r_{i,1}, r_{i,2}, \ldots, r_{i,n})
\end{equation}

\begin{center}
\begin{table}
\caption{List of notations used in this paper.}
\centering
\begin{tabular}{|c | l |} 
 \hline
 Notation & Description\\ 
 \hline\hline
 $U$ & Set of users\\
 \hline
 $u_i$ & An individual user\\
 \hline
 $m$ & Number of users\\
 \hline
 $O$ & Set of items or options\\
 \hline
 $o_j$ & An item or option\\
 \hline
 $n$ & Number of options\\
 \hline
 $R$ & User-option rating matrix\\
 \hline
 $r_{i,j}$ & User $u_i$ rating of option $o_j$\\
 \hline
$\boldsymbol{u}_i$ & Profile vector of user $u_i$\\
 \hline
 $g$ & group of users\\
 \hline
$\boldsymbol{g}$ & Group $g$'s profile vector\\
 \hline
 $\boldsymbol{g}^{AG}$ & Group $g$'s profile vector constructed using the $AG$ aggregation strategy\\
 \hline
 $f^{AG}$ & A function for calculating group profile using $AG$ aggregation strategy\\
 \hline
 $c(g)$ & Actual choice of $g$\\
 \hline
 $c^{*} (g)$ & Predicted choice of group $g$\\
 \hline
 $\mathcal{G}$ & A set of tuples each of which consists of a group profile and group choice\\
 \hline
 $G_{\text{train}}$ & A subset $\mathcal{G}$ used for training the model\\
 \hline
 $G_{\text{test}}$ & A subset $\mathcal{G}$ used for testing the model\\
 \hline
 $\sigma(\cdot)$ & A permutation function\\
\hline

\end{tabular}
\label{tab:notation}
\end{table}
\end{center}

Table~\ref{tab:User_Rating} shows an example of a rating matrix. In this example, four users are present, $u_1$, $u_2$, $u_3$, $u_4$, and ten options are listed: $o_1, \ldots, o_{10}$. Each number in the table indicates the corresponding user-option rating. In this example, all the possible users' ratings are known: the matrix is complete. Throughout the remainder of the paper, this user-option ratings matrix example (Table~\ref{tab:User_Rating}) will be utilized to illustrate the proposed techniques.

\begin{center}
\begin{table}[ht]
\caption{Example of a rating matrix of users $u_1$, $u_2$, $u_3$, and $u_4$, 10 options $\{o_1, ..., o_{10}\}$ and their ratings, given in a 1-10 rating scale.}
\centering
\begin{tabular}{|c | c c c c c c c c c c |} 
 \hline
 Member & $o_1$ & $o_2$ & $o_3$ & $o_4$ & $o_5$ & $o_6$ & $o_7$ & $o_8$ & $o_9$ & $o_{10}$\\ 
 \hline\hline
 ${u}_1$ & 6 & 9 & 4 & 8 & 5 & 2 & 7 & 1 & 10 & 3\\
 \hline
 ${u}_2$ & 7 & 6 & 4 & 1 & 2 & 10 & 3 & 8 & 9 & 5\\
 \hline
 ${u}_3$ & 1 & 10 & 3 & 5 & 9 & 6 & 8 & 7 & 2 & 4\\
 \hline
 ${u}_4$ & 6 & 8 & 3 & 9 & 1 & 5 & 7 & 2 & 10 & 4\\
 \hline
\end{tabular}
\label{tab:User_Rating}
\end{table}
\end{center}

A group profile is an aggregated representation of the preferences of the group members. It is a real vector of the same dimensionality as the group members' profiles and the value of each entry indicates the ``importance'', or ``score'' of the corresponding option for the group. A group profile can be obtained by applying a preference aggregation strategy to the group members' profiles. Let $g = \{u_1, u_2,..., u_{|g|}\} \subset U$ be a group. We denote with $\boldsymbol{g}$ a group profile of $g$. We will also  denote with $\boldsymbol{g}^{AG}$ a profile of $g$ when we want to make evident the preference aggregation strategy $AG$ used to generate the group profile:
\begin{equation} \label{eq1}
\begin{split}
 &\boldsymbol{g}^{AG} = f^{AG}(\boldsymbol{u}_1, \ldots, \boldsymbol{u}_{|g|})\\
 & \boldsymbol{g}^{AG} = (r_{g,1}, \ldots,  r_{g,n})
\end{split}
\end{equation}
where $f^{AG}$ is a preference aggregation strategy function; $r_{g,j}$, the group score for option $j$, is a non negative real number. In Section \ref{sec:aggr_strat}, we have introduced the Average (AVE), Multiplicative (MULT), Least Misery (LM), and Copeland Rule (COPE) preference aggregation strategies. In AVE, the group $g$ score for option $j$, i.e., $r_{g, j}$ is the average of group members' ratings $\{r_{u_{1}, j}, \ldots, r_{u_{|g|}, j}\}$ for that option $j$. In MULT, the group score for an option is the multiplication of group members' ratings for that option, and in LM the group score is the minimum rating of the group members' ratings for that option. We now define the Copeland Rule-based profile, the original Stasser group profile~\cite{stasser1999}, and a generalization of that profile.

\textbf{Copeland Rule (COPE).} For calculating the group profile of $g$ according to the Copeland rule, one needs to compute a real $n \times n$ matrix $M^g = (m_{i,j})$ where each option in $O$ corresponds to a column and a row in this matrix. Each entry of this matrix is computed with the formula:

\[ m_{i,j} =
        \begin{cases}
            1   \quad \text{if } & |\{u\in g \hspace{1ex} | \hspace{1ex} r_{u, i} < r_{u, j} \}| > \\
            & |\{u  \in g \hspace{1ex} | \hspace{1ex} r_{u, i} > r_{u, j} \}|\\
            
            0   \quad \text{if } & |\{u\in g \hspace{1ex} | \hspace{1ex} r_{u, i} < r_{u, j} \}| = \\
            & |\{u \in g \hspace{1ex} | \hspace{1ex} r_{u, i} > r_{u, j} \}|\\
            
            -1   \quad \text{if } & |\{u\in g \hspace{1ex} | \hspace{1ex} r_{u, i} < r_{u, j} \}| < \\
            & |\{ u \in g \hspace{1ex} | \hspace{1ex} r_{u, i} > r_{u, j} \}|\\
        \end{cases}
    \]
Then the COPE group score for option $j$ is equal to the sum of the values in column $j$ of the  $M^g$ matrix:
\begin{equation}
    r_{g,j} = \sum_{i=1}^{n} m_{i,j}
\end{equation}

\textbf{Stasser group profile (SDS1) and generalised version (SDS3).} As we have discussed in Section \ref{sec:sds}, the Stasser \cite{stasser1999} model for predicting a group choice is based on a group representation that counts, for each option, the number of group members that prefer that option (to the others). We call this preference aggregation strategy SDS1. We also consider a straightforward generalization of this approach, named SDS3 where the group's score for an option is obtained by counting the number of times that option is among the top three preferred options of the group members. Similarly can be defined SDS2, SDS4, and SDSn. However, in order to simplify the analysis of the results, in the following we will only consider SDS3.

\subsection{Unknown User Ratings and Normalization} 
\label{sec:unknownratings}

It is worth noting that some entries of the rating matrix $R$ might be unknown, i.e., one or more group members may have not rated an option. In such cases, which are not considered in the original SDS formulation, the group score for that option is computed by using only the available group members' ratings for the option. When none of the group members has rated an option, one can label the group score of that option as ``unknown''.

Finally, after applying any preference aggregation strategy the group profile is normalized so that the sum of its entries is $1$. Hence if $\boldsymbol{g} = ({r}_{g,1}, \ldots, {r}_{g,n})$ is a group profile obtained by any preference aggregation strategy, then after normalization, the entries of this vector are replaced with:

\begin{equation}
\label{eq:normalization}
    {r}_{g,j} := \frac{r_{g,j}}{\sum_{ k = 1}^{n} r_{g,k}}
\end{equation}
For instance, Table \ref{tab:group_Scores} shows group profiles of the group $g=\{u_1, u_2, u_3, u_4 \}$ whose individual ratings are shown in Table~\ref{tab:User_Rating}. In this example, $o_2$ is the preferred option of $u_3$, since $r_{3,2} = 10$, $o_6$ is the preferred option of $u_2$, for the same reason, and $o_9$ is the preferred option of $u_1$ and $u_4$. Then, for example, in the SDS1-based group profile, before normalization, the group score for options $o_2$ and $o_6$ is $1$, and for $o_9$ is $2$. After normalization, the group scores of $o_2$ and $o_6$ are $0.25$ and the group score of $o_9$ is $0.5$, and the group profile is $\boldsymbol{g}^{SDS1} = (0, 0.25, 0, 0, 0, 0.25, 0, 0, 0.5, 0)$.

\begin{table}[!htbp]
\centering
\caption{Normalized group profiles calculated by using the preference aggregation strategies AVE, MULT, LM, SDS1, SDS3, and COPE. Users' ratings are as in Table~\ref{tab:User_Rating} and the group is $g=\{u_1, u_2, u_3, u_4\}$. Each number in the rows of this table is the group score for the column item, employing the preference aggregation strategy indicated in the first column of the row.}
\begin{tabular}{|c | c c c c c c c c c c |} 
 \hline
 group profile & $o_1$ & $o_2$ & $o_3$ & $o_4$ & $o_5$ & $o_6$ & $o_7$ & $o_8$ & $o_9$ & $o_{10}$\\ 
 \hline\hline
 $\boldsymbol{g}^{AVE}$ & 0.09 & 0.15 & 0.06 & 0.1 & 0.07 & 0.1 & 0.11 & 0.08 & 0.14 & 0.07\\
 \hline
 $\boldsymbol{g}^{MULT}$ & 0.02 & 0.47 & 0.01 & 0.03 & 0.009 & 0.06 & 0.12 & 0.01 & 0.19 & 0.02\\
 \hline
 $\boldsymbol{g}^{LM}$ & 0.04 & 0.26 & 0.13 & 0.04 & 0.04 & 0.08 & 0.13 & 0.04 & 0.08 & 0.13\\
 \hline
 $\boldsymbol{g}^{SDS1}$ & 0 & 0.25 & 0 & 0 & 0 & 0.25 & 0 & 0 & 0.5 & 0\\
 \hline
 $\boldsymbol{g}^{SDS3}$ & 0 & 0.25 & 0 & 0.16 & 0.08 & 0.08 & 0.08 & 0.08 & 0.08 & 0.25\\
 \hline
  $\boldsymbol{g}^{COPE}$ & 0.08 & 0.2 & 0 & 0.11 & 0.04 & 0.11 & 0.15 & 0.042 & 0.21 & 0.02\\
 \hline
\end{tabular}
    \label{tab:group_Scores}
\end{table}

\subsection{Predicting Group Choice}
\label{sec:pred_group_choice}

\subsubsection{Preference Aggregation based Choice Prediction - PACP}

A preference aggregation strategy can be used to predict a group choice: the option that, according to the selected preference aggregation strategy, has the largest score. This approach is here considered as a baseline method. Hence, if $\boldsymbol{g}^{AG} = (r_{g,1}, \ldots, r_{g,n})$ is the $g$'s group profile calculated by the preference aggregation strategy $AG$, then the {\it Preference aggregation strategy based Choice Prediction} method (PACP) predicts that the group with profile $\boldsymbol{g}^{AG}$ will choose the option:
\begin{equation}
    c^{*} (g) = \arg \max_{j \in O} \{ r_{g,j} \}
\end{equation}
We use the star notation, $c^{*} (g)$, to indicate the predicted choice, while the actual choice is denoted with $c(g)$. We also stress that PACP operates on a group profile built by using a preference aggregation strategy, hence, when it is needed, we will use a distinct notation, such as PACP-AVE, to explicitly indicate the preference aggregation strategy (AVE) used in the prediction\footnote{We note that when computing choice predictions in our experiments we have also used Borda count and Most Pleasure preference aggregation strategies. However, we do not discuss these prediction methods because PACP with the Borda aggregation strategy has a performance very similar to that obtained by the Average strategy. Besides, the performance of Most Pleasure was consistently worse than any other strategy.}.

\subsubsection{Learning-based Choice Prediction - LCP}

PACP  is rigidly employing a preference aggregation strategy when predicting a group choice. Motivated by SDS, we conjecture that the choice can be better predicted by leveraging the analysis of patterns of user preferences encoded in a group profile, which is constructed by using a preference aggregation strategy. Hence, after the construction of the group profile, based on the group members' ratings and a selected preference aggregation strategy, we use a machine learning classifier to predict the actual group choice.

In order to implement this idea, given a set of groups $G$ we need for each group $g \in G$: the group members' individual preferences to be aggregated in a group profile $\boldsymbol{g}$, and the observed choice made by the group, i.e., $c(g)$. Then, this data set $\mathcal{G} = \{(\boldsymbol{g}, c(g)): g \in G\}$ is given as input to a Machine Learning algorithm, which in this paper is Multinomial Logistic Regression, that generates a prediction $c^*(g)$ of the actual group choice $c(g)$. Then, the group choice is a class variable, taking value in the set of all the available options $O$. Thus, the group choice prediction problem is translated into a classification problem: given a training set consisting of group profiles in $\mathcal{G_{\text{train}}}$, where the group choice is known, after having trained the ML classifier on that set, the classifier can predict the choice $c^*(g')$ of a group $g'$ in a test set $\mathcal{G_{\text{test}}}$. 

Similarly to what was said for PACP, we stress that LCP operates on a set of group profiles built by using a preference aggregation strategy, consequently, when needed, we will use a distinct notation, such as LCP-AVE, to explicitly indicate the used preference aggregation strategy (AVE).

\section{Data Augmentation with Synthetic Group Profiles}
\label{sec:data_augm}

In this section, we face the data scarcity problem by presenting two data augmentation methods. LCP group choice prediction is trained on a data set of observed groups, along with their members' preferences and the final choice of the group. Clearly, a larger training data set will give to the model more information about the target choice function to be learned. However, often the training data set is small, as the dataset used in our experiments, which contains only 79 groups. This is a limiting factor for any Machine Learning model. While the minimum number of required sample size depends on different factors, in~\cite{lakshmanan2020machine} it is claimed that a reasonable minimum number of instances for a classification problem with $c$ classes, where instances are described by $f$ features, is $ 10 f c$. So, for instance, in the experiments conducted in this paper $f = c = 10$, and this means that a minimum of 1000 groups would be needed, while instead, our data set is one order of magnitude smaller.

To address this problem, we make some assumptions about the functional relationship between the group profile and the group choice. These assumptions are then motivating the creation of synthetic groups that were not actually observed but should be observed, given the assumptions. 
The first assumption on the choice function is that in a group where all the group members prefer the same option, that option would be chosen by the group. Hence, even if groups, where all the group members prefer a single option, were not observed, we introduce in the training set synthetic groups of this type, and we set as group choice the option that is preferred by all. 
The second assumption is that the order of the options in the vector representation of the group profile is not relevant for the choice function. Instead, it is assumed that only the relative scores of options in the vector representation of the group profile impact the choice function. Hence, assume that we have observed a group $g$ that chose the first option. Then, if a group $g'$ has a group profile that is a permutation of the profile of group $g$, i.e., the same scores but in a different order, then the choice of $g'$ must be the option that by the permutation corresponds to the first option in group $g$ profile.

We are therefore following a data augmentation approach and we expand the training data set by adding synthetic groups (group profile and group choice) that could improve the choice prediction function on the test set. We generate two sets of synthetic group profiles, \textit{Winners} and \textit{Permutations}, which are presented below. 

We note that the idea of using synthetic variations of the training examples, such as those introduced and presented below, comes from similar data augmentation methods which are often used in Machine Learning. Data augmentation is, in fact, a technique to increase the diversity and the coverage of the training set by applying random, however realistic, transformations when the data set is not large enough for a trained model to generalize well~\cite{antoniou2017data,shorten2019survey}.

\paragraph{Winners}

The first set of synthetic group profiles, is called {\it Winners} and contains one group profile  $\boldsymbol{g}_{j} = (r_{g_j,1}, \dots, r_{g_j,n})$ for each option $j \in O$. A group in the \textit{Winners} set has profile scores concentrated on a single option which is also assumed to be the group choice:
\[ r_{g_j,k} =
        \begin{cases}
            1  & \quad \text{if } k = j\\
            0  & \quad \text{otherwise}
        \end{cases}
\]
and 
\begin{equation}
    c(g_j) = j
\end{equation}
Hence, when LCP is trained on the set of group profiles $G_{train}$ we add to that training set the group profiles in the \textit{Winners} set:
\begin{equation}
    G_{train}^{Win} = G_{train} \cup \{(\boldsymbol{g}_{j}, j) | j \in O\}
\end{equation}

Table \ref{tab:winner} shows three examples of Winner group profiles when $|O| = 10$. The motivation for adding such, possibly missing, observations is that they comply with a fundamental axiom of social choice: if all the group members prefer the same option, this must be the group choice. The addition of these synthetic profiles could ease the training of the Machine Learning algorithm that predicts the group choice, as it explicitly adds knowledge that the predictive model might not be able to extract from the available data. 
\begin{center}
\begin{table}
\caption{Winners group profiles. In each profile, there is only one option with a group score equal to 1, while the remaining scores are zero. The option with group score 1 is the group choice.}
\label{tab:winner}
\centering
\begin{tabular}{|c | c c c c c c c c c c | c|} 
 \hline
 Group & $o_1$ & $o_2$ & $o_3$ & $o_4$ & $o_5$ & $o_6$ & $o_7$ & $o_8$ & $o_9$ & $o_{10}$ & Group choice\\ 
 \hline\hline
 $\boldsymbol{g}_1$ & 1 & 0 & 0 & 0 & 0 & 0 & 0 & 0 & 0 & 0 & $o_1$\\
 \hline
 $\boldsymbol{g}_2$  & 0 & 1 & 0 & 0 & 0 & 0 & 0 & 0 & 0 & 0 & $o_2$\\
 \hline
 
 &&&&&& $\ddots$ &&&&& \\
 
 \hline
 $\boldsymbol{g}_{10}$  & 0 & 0 & 0 & 0 & 0 & 0 & 0 & 0 & 0 & 1 & $o_{10}$\\
 \hline
\end{tabular}
\end{table}
\end{center}

\paragraph{Permutations}

The second set of synthetic group profiles is based on an assumption that is at the base of the SDS theory: the group choice should not depend on the option itself, but on the relative scores of options in the group profile. Hence, the \textit{Permutations} profiles are generated by cloning existing profiles and rearranging the order of the options, their scores, and the group choice accordingly. In mathematics, a permutation of a list is a change in the ordering of the elements of the list. For instance, $(1, 3, 2)$, $(2, 1, 3)$, $(2, 3, 1)$, and $(3, 1, 2)$ are all the permutations of the ordered list $(1, 2, 3)$. Given for instance a group profile $\boldsymbol{g} = (r_{g,1}, r_{g,2}, r_{g,3})$ with 3 options, a permutation of this profile could be $\boldsymbol{g}' = (r_{g,2}, r_{g,1}, r_{g,3})$, where the first and the second scores are swapped. Let us now assume that the group profile $\boldsymbol{g}$ belongs to a group $g$ that chose the first option. If this choice is determined by the relative values of the scores  $(r_{g,1}, r_{g,2}, r_{g,3})$, then one can assume that for a group with profile $\boldsymbol{g}' = (r_{g,2}, r_{g,1}, r_{g,3})$ the choice will be the second option, which has exactly the same score as the first option in the profile $\boldsymbol{g}$. To give another example, consider the group profiles in Table~\ref{tab:winner}. The profiles $\boldsymbol{g}_2, \ldots, \boldsymbol{g}_{10}$ are permutations of the first profile $\boldsymbol{g}_1$, and if in the first profile, the group choice is the first option, it is evident that in the second profile, the option chosen by the group must be the second, and so on.

We now give a formal description of the data augmentation approach based on the permutation of group profiles in a data set. Let $G_{train}$ be the available training set for LCP, i.e., it contains pairs $\big(\boldsymbol{g}, c(g)\big)$ composed by a group profile $\boldsymbol{g}$ and the (observed) group choice $c(g)$. Let $\sigma$ be a permutation of $O$, i.e., a rearrangement of the $n = |O|$ options, $\sigma: \{1,...,n\} \longrightarrow \{1,...,n\}$. By using a group profile $\big(\boldsymbol{g}, c(g)\big) \in G_{train}$, and a permutation $\sigma$, we create a new group profile $\sigma(\boldsymbol{g})$ and the corresponding group choice $c\big(\sigma(\boldsymbol{g})\big)$ as follows:
\begin{equation}
\begin{split}
    & \sigma(\boldsymbol{g})  = (r_{g,\sigma(1)}, r_{g,\sigma(2)},..., r_{g,\sigma(n)})\\
    & c(\sigma(\boldsymbol{g})) = \sigma(c(g))\\
\end{split}
\end{equation}

Table \ref{tab:permutation} shows a group profile and two examples of permuted group profiles. $o_2$ is the actual group choice of the group with profile $\boldsymbol{g}$: it is the option with the largest group score. $\sigma_1$ is a permutation that reorders all the options and maps $o_2$ to $o_1$, while $\sigma_2$ is a permutation that maps $o_2$ to $o_3$. If the group $g$, with the particular pattern of group scores as in $\boldsymbol{g}$, made the choice $o_2$, then if the scores are only rearranged, hence their relative values are not changed, the choice of the group with profile $\sigma_1(\boldsymbol{g})$ now should be the option corresponding to the option chosen in the original group, hence it must be $o_1$.  Similarly, one can also comment on $\sigma_2$.

\begin{center}
\begin{table}[ht]
\caption{Examples of synthetic data constructed using the Permutation augmentation method. In this table, $\boldsymbol{g}$ is the group profile, and $\sigma_1$ and $\sigma_2$ are the group profiles constructed using the permutation augmentation method.}
\label{tab:permutation}
\centering
\begin{tabular}{|c c c c c c c c c c c c|} 
 \hline
 Group Profile & $o_1$ & $o_2$ & $o_3$ & $o_4$ & $o_5$ & $o_6$ & $o_7$ & $o_8$ & $o_9$ & $o_{10}$ & Group choice\\ 
 \hline\hline
 $\boldsymbol{g}$ & 0.09 & \textbf{0.15} & 0.06 & 0.1 & 0.07 & 0.1 & 0.11 & 0.08 & 0.14 & 0.07 & $o_2$\\
 \hline
 $\sigma_1(\boldsymbol{g})$ & \textbf{0.15} & 0.1 & 0.07 & 0.11 & 0.1 & 0.08 & 0.07 & 0.14 & 0.09  & 0.06 & $o_1$\\
 \hline
 $\sigma_2(\boldsymbol{g})$  & 0.11 & 0.08 & \textbf{0.15} & 0.07 & 0.07 & 0.1 & 0.14 & 0.09 & 0.06 & 0.1 & $o_3$\\
 \hline
\end{tabular}
\end{table}
\end{center}

So, given a training set $G_{train}$  of group profiles and corresponding choices, we sample with repetition from this training set and obtain a new set (with repetitions) of group profiles and corresponding choices $\{(\boldsymbol{g}_1, c(g_1)), \ldots,  (\boldsymbol{g}_N, c(g_N))\}$, where $\big(\boldsymbol{g}_l, c(g_l)\big) \in G_{train}$, $l= 1, \ldots, N$. We then select a permutation $\sigma_l$ of $O$ for each profile $\big(\boldsymbol{g}_l, c(g_l)\big)$, $l= 1, \ldots, N$, and we generate a new permuted group profile $\big(\sigma_l(\boldsymbol{g}_l), \sigma(c(\boldsymbol{g}_l)\big)$. We add these permuted profiles to the original training set: 
\begin{equation}
    G_{train}^{Perm} = G_{train} \cup \{ (\sigma_1(\boldsymbol{g}_1), \sigma_1(c(g_1))), \ldots, (\sigma_N(\boldsymbol{g}_N), \sigma_N(c(g_N))) \}
\end{equation}
Algorithm \ref{alg:perm} shows the exact procedure that we have designed for constructing new profiles using the Permutation data augmentation method. 
\begin{algorithm}
\caption{Augmentation of the training set with permutations of available group profiles}
\label{alg:perm}
\begin{algorithmic}
    \State Given a set of groups $G \subset \wp^{U}$ and their profiles $G_{train}$ = \{$(\boldsymbol{g}, c(g)): g \in G$\}
    \State Given a probability distribution over the choice set $p = (p_1, p_2,...,p_n)$
    \State $N$ is the number of group permutations to generate
    \State $G_{train}^{Perm} := G_{train}$ is the new training set to build
    \For{N times}
            \State Sample a choice $j \in O$ with the probability distribution $p$
            \State Sample a group $g \in G$ where $c(g) \neq j$
            \State Generate a permutation $\sigma$, s.t. $\sigma(c(g)) = j$ 
            \State Add $(\sigma(\boldsymbol{g}), j)$ to $G_{train}^{Perm}$
    \EndFor
        \State \textbf{return} $G_{train}^{Perm}$
    
\end{algorithmic}
\end{algorithm}
In this algorithm, $N$ permuted group profiles are generated by making sure that a target distribution of group choices $p$ is preserved in the synthetic data. This distribution $p$, in our experiments, is the observed distribution of the choices in the training data set, hence, by adding the permuted profiles, we do not change the distribution of the choices in the training set. The rationale of this is related to a typical bias of Machine Learning algorithms, of being influenced by the class distribution in the training set.
Finally, we note that the parameter $N$ should be selected case by case, depending on the available data and the complexity of the choice prediction problem. In our experiments, we have added to the training set (of 60 groups) 1200 permutations (more details are given in the next section).

%% file: 05_Experiments.tex
\section{Experimental Evaluation}
\label{sec:experiments}

\subsection{Dataset}
\label{sec:dataset}

The groups' observational data, which is used in this paper, was collected in a user study focused on the travel and tourism domain. The study was implemented in two rounds at several universities in Europe. Both implementations followed the same three-phase structure~\cite{GROUPDM18,delic2018social,GROUP_DM16}. 
In the first phase of the user study, the participants' explicit preferences, i.e., either ratings or rankings for ten pre-selected destinations (options) were collected. The two rounds differed in the pre-selected destinations, and in the way participants expressed their preferences about them. In the first round, the destinations were ten large European cities and the participants {\it ranked} them. Conversely, in the second round, the destinations were chosen to fit the general preferences of certain traveler types identified in the tourism  literature~\cite{Yiannakis1992,GIBSON02,NEIDHARDT2014,gretzel2004,moscardo1996}, and the participants {\it rated} them on a ten-point scale (1 - not attractive, 10 - highly attractive). The rationale for changing the destination set was to increase their diversity and consequently also the preferences of the group members. 

In the second phase of the study, the participants were asked to form groups freely, with the only restriction that the group size should not exceed five members. Each user participated in only one group. The rationale of the group size constraint was only to focus the acquired data on the typical scenario of small groups and to avoid collecting data from a smaller number of larger groups. Then, the participants joined their respective groups and started a face-to-face discussion aimed at selecting, from the pre-defined set, a destination that they as a group would like to visit together. It is worth noting, that in the face-to-face discussion, the group members did not have access to the ratings or rankings, which they previously, individually, assigned to the options. Hence, these users, while making a choice in their group, might not have recalled precisely the expressed preferences, and the group discussion could have brought the group to choices that are not well justified by the pre-discussion individual preferences.

Finally, in the third phase, the participants filled in a post-questionnaire, where they indicated which destination the group selected, i.e., the group choice. The observational study resulted in two data sets: DSI (200 participants in 55 groups), and a smaller set DSII (82 participants in 24 groups)\footnote{The combined datasets are available at \url{https://github.com/amradelic/Tourism-dataset}.}. 

In order to compute the group profiles, all the considered strategies require the group members' ratings as input. As mentioned, the two data sets, DSI and DSII, were different in terms of group members' individual preferences, i.e., rankings in DSI and ratings in DSII. Hence, to derive ratings of the options from users' ranked lists in DSI, we assign the maximum score (10) to the first option in a user's ranked list, 9 to the second option, etc.

Finally, in~\cite{masthoff2006pursuit} it was shown that in GRSs it may be beneficial to alter the original ratings of the users in order to amplify the difference between highly rated items and lower-rated ones. Hence, in order to implement this idea, we have replaced the original ratings with the square of them.

\subsection{Evaluation setting}

In order to evaluate the performance of the proposed group choice prediction approach, namely LCP, we have compared it to the baseline PACP method. To train LCP, we use the Multinomial Logistic Regression classifier~\cite{venables2013modern}. 
We have also tested the performance of other classifiers, such as Linear Discriminant Analysis \cite{venables2013modern,ripley2007pattern} and Support Vector Machines with a linear kernel function \cite{chang2011libsvm}. However, Multinomial Logistic Regression outperformed these models, very likely because of its simplicity and the limited size of our data set. For that reason, we only show the performance of LPC when the Multinomial Logistic Regression classifier is used.

Four-fold cross-validation is employed to estimate LCP and PACP performance. Four folds are selected due to the size of the data set (i.e., 79 groups altogether). In fact, having more than four folds would produce folds containing less than 15 instances. As a measure of performance, we report the average accuracy of the predicted group choices (i.e., the number of correct predictions over the number of all predictions). Since our data set is small and the estimated accuracy of LCP can depend on the particular four folds used in the cross-validation, we iterated the whole procedure ten times and we reported the average accuracy of these ten iterations. In other words, in each iteration, we calculated the accuracy using the standard cross-validation method, and then, we reported the average of these ten accuracy values. We note that in order to have a fair comparison, we have used the same foldings, in the ten repetitions, to evaluate all the considered models and their variants. We have implemented our models in Python and have used the scikit-learn library. To tune the hyperparameters, we have utilized the Grid Search function available in the scikit-learn. We employed various solvers, namely 'newton-cg', 'liblinear', 'lbfgs', 'sag', and 'saga'. The regularization term interval was set to $[0.1, 50]$, with a step size of 0.1. 

PACP baseline predicts the option with the highest score as the group choice. However, there may be group profiles in which more than one option has the same largest score. In this case, PACP selects one of these alternatives randomly. Hence, also for the evaluation of PACP, we repeated the four-fold validation procedure ten times and the final result is the average accuracy of these ten repetitions. In each of these ten repetitions, we used the same foldings and cross-validation method, but PACP does not require any learning phase, the choice prediction on a test group is only based on the group profile data. 

As previously stated, we created our dataset by merging two distinct datasets, namely DSI and DSII. In order to assess the influence of this combination, we proceeded to replicate our experiment on the larger dataset alone. This experiment also aimed to examine whether the proposed ML approach is robust even when combining different datasets. 

In our data set the users have rated all of the items. However, in real-world scenarios, it is common for some or even most of the alternatives to be unrated by users. Thus, to assess the \textit{robustness} of our approach and compare it to the baseline, we have performed an additional experiment. In this experiment, we randomly eliminated certain user-option ratings, before creating the group profiles. By varying the probability to eliminate a rating, we create a collection of new data sets of group profiles. We then predict the group choices by considering, independently, these data sets. Hence we test our group choice prediction method relying on sparse matrices of user-option ratings. We followed the previously mentioned procedure for our experiment. For each of the user-option matrices, we generated by removing certain individual ratings, we employed a four-fold cross-validation approach. We repeated this process ten times and calculated the average accuracy of choice predictions for both PACP and LCP.

We have also compared the prediction performance of LPC with humans' ability in predicting group choice by only knowing the individual group members' ratings. To do so, we have developed a user interface in which the participants could check the individual ratings of the group members in our dataset and predict the group choice. More details on this GUI and the experiment are given in the results section. 

We have finally evaluated the performance of LPC when the Winners and Permutations data augmentation approaches are used. Similarly to what is described above, when the training set is augmented with \textit{Permutations} in order to avoid the possibility of having training sets of different quality, we repeat the four-fold cross-validation procedure ten times; by generating each time a different set of synthetic group profiles. We stress that \textit{Winners} and \textit{Permutations} profiles are only added to the training set, while the test set contains only genuine group profiles: the test set was not changed, in any fold nor repetition. 
The number of Winner profiles that are added to the training set is 10, as this is the number of options in our prediction task. The number of Permutations added is 1200. This number was optimized with cross-validation.

In summary, considering the two models for group choice prediction, namely, PACP and LCP, the diverse types of group profiles produced by the considered preference aggregation strategies (AVE, MULT, LM, COPE, SDS1, and SDS3), and the two data augmentation approaches (\textit{Winners} and \textit{Permutations}) we have generated $3 * 6 = 18$ variants of LCP and $6$ variants of PACP. Table \ref{tab:ac} reports the names of these variants and their characteristics. For instance, the aggregation strategy AVE is used to generate group profiles that are considered in: LCP-AVE, LCP-AVE-W (\textit{Winners}), LCP-AVE-P (\textit{Permutations}) and PACP-AVE. 

\begin{center}
\begin{table}[ht]
\caption{List of the choice prediction model variants that are considered in the evaluation.}
\label{tab:my_label}
\centering
\begin{tabular}{| c | c c c c |} 
 \hline
 Pref. Aggr. Strat. & Winners & Permutations & PACP & LCP \\ 
 \hline\hline
 \multirow{3}{*}{AVE} & - & - & PACP-AVE & LCP-AVE \\ \cline{2-5}
  & \checkmark & - & - & LCP-AVE-W \\ \cline{2-5}
  & - & \checkmark & - & LCP-AVE-P \\ 
 \hline\hline
 \multirow{3}{*}{MULT} & - & - & PACP-MULT & LCP-MULT \\ \cline{2-5}
   & \checkmark & - & - & LCP-MULT-W \\ \cline{2-5}
  & - & \checkmark & - & LCP-MULT-P \\ 
 \hline\hline
 \multirow{3}{*}{LM} & - & - & PACP-LM & LCP-LM \\ \cline{2-5}
  & \checkmark & - & - & LCP-LM-W \\ \cline{2-5}
  & - & \checkmark & - & LCP-LM-P \\ 
 \hline\hline
 \multirow{3}{*}{SDS1}  & - & - & PACP-SDS1 & LCP-SDS1 \\ \cline{2-5}
  & \checkmark & - & - & LCP-SDS1-W \\ \cline{2-5}
  & - & \checkmark & - & LCP-SDS1-P \\ 
 \hline\hline
 \multirow{3}{*}{SDS3}  & - & - & PACP-SDS3 & LCP-SDS3 \\ \cline{2-5}
   & \checkmark & - & - & LCP-SDS3-W \\  \cline{2-5}
  & - & \checkmark & - & LCP-SDS3-P \\ 
 \hline\hline
 \multirow{3}{*}{COPE}  & - & - & PACP-COPE & LCP-COPE \\ \cline{2-5}
  & \checkmark & - & - & LCP-COPE-W \\ \cline{2-5}
  & - & \checkmark & - & LCP-COPE-P \\ 
 \hline
\end{tabular}
\label{tab:ac}
\end{table}
\end{center}

%% file: 06_Results.tex
\section{Results}
\label{sec:results}

We address the first hypothesis stated in Section~\ref{sec:hyp} by evaluating the predictive capability of the proposed LPC model in predicting the group choice. We assess also the robustness of the proposed models in dealing with unknown user ratings, and we present and discuss the findings from our user study regarding the human ability to predict group choice. Then we move to the second hypothesis and we assess the effectiveness of the proposed data augmentation method.

\subsection{Predictive Capability of LCP}
We start by examining the validity of our first hypothesis: \textit{by using a proper machine learning model, trained on a data set of group choices, it is possible to effectively predict the choice of a group, by using only the individual preferences of the group members, encoded in a group profile.}

\begin{figure}[!htbp]
    \centering
    \includegraphics[width=0.9\linewidth]{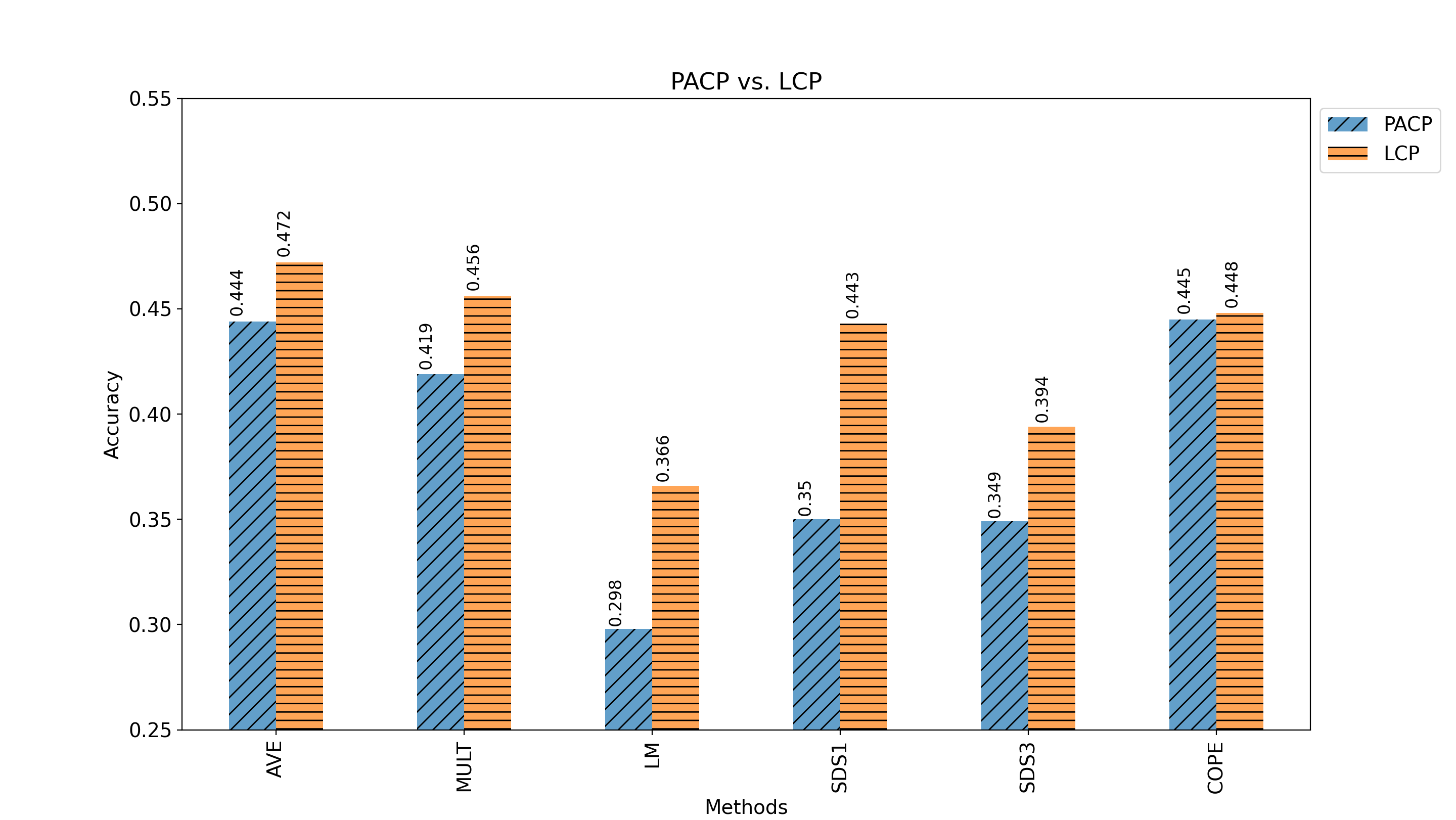}
    \caption{Comparison of the accuracy of the LCP and PACP variants. As indicated in this figure, LCP-AVE outperforms the other LCP and PACP variants. Additionally, for all variants, LCP-* outperforms the corresponding variant of PACP.}
    \label{Square_Standard_Strategies_font}
\end{figure}

Figure \ref{Square_Standard_Strategies_font} shows the performance (choice prediction accuracy) of the considered LCP and PACP variants. The benefit of using  LCP  in comparison to PACP is clear: for all the considered group profile generation methods (AVE, MULT, LM, SDS1, SDS3, and COPE), LCP  achieves a better performance than the corresponding baseline PACP variant. We note that LCP-AVE, i.e., LCP when the group profile is constructed with the AVE aggregation strategy, outperforms all the other LCP variants, i.e., those using group profiles produced by the alternative preference aggregation strategies. So, by using the AVE preference aggregation strategy to build the group profiles, LPC can reach a good performance of almost 50\% accuracy. It is also interesting to note that even if SDS1 and SDS3 preference aggregation strategies are not among the best, there is a great benefit of using the LCP learning approach when these strategies are used for building the group profile. This confirms Stasser's intuition in the SDS theory that the choice function can be learned from group choice data. 

\subsection{Considering DSI and DSII Separately.}
\label{sec:combinig_data}

As mentioned in Section \ref{sec:dataset}, the data set that we use in our experiments is the union of two data sets (DSI and DSII) collected in two implementations of the \cite{GROUPDM18,delic2018social,GROUP_DM16} user study. These two implementations are different w.r.t. the options considered by the participants and the type of collected individual preferences (ratings vs ranking). Merging these two data sets for creating a single one is also motivated by the assumption introduced by Stasser \cite{stasser1999} and mentioned above: \textit{the specific pattern of group scores in the group profile and not the options mostly determines the group choice}. 

In order to assess the assumption that by using the merged data set, instead of the two independently, would not negatively impact on the accuracy of the LCP models, we repeated our experiment on DSI, the larger one, which contains 55 groups. The second small data set contains only 24 groups, which makes it inappropriate for properly assessing the quality of both LCP and PACP.

By comparing the accuracy of the LCP variants in the merged data set (DSI+DSII) with the corresponding variants in DSI, we have discovered that the prediction accuracy barely differs. For instance, the accuracy of LCP-AVE (LCP-MULT) on the merged data set is only 0.003 (0.032) larger than the accuracy of LCP-AVE (LCP-MULT) on DSI. 

This result supports the above-mentioned assumption that \textit{the specific distribution of the group scores that are in the group profile and not the options determines the group choice}. In fact, the merged data set, which we use in our analysis, is the combination of group profiles related to two different choice tasks: the common aspect is only the same number of options. Hence, this analysis confirms the validity of using the merged data set, derived from two independent implementations of the destination selection task. It also shows that one can obtain a benefit by merging data coming from somewhat different group decision tasks.

\subsection{Dealing with Unknown User Ratings}
\label{sec:sparse_ratings}

The user-option rating matrix used in our experiments is dense: all the users have rated all the options. However, in many practical situations, this may not be the case: a group member may express his/her preferences only for a subset of the available options. To evaluate the ability of PACP and LCP to deal with these cases, i.e., when the user-option rating matrix is sparse and there are unknown user ratings, we produced a collection of new sub-datasets by discarding in each sub-dataset a proportion of the group members' ratings. To do so, in each sub-dataset user ratings were, one by one, independently removed with a given probability $p$. We have considered probability values between $0$ and $0.6$ (with $0.01$ step) to produce a collection of sub-datasets.

As mentioned in Section \ref{sec:unknownratings}, to calculate the group score for an option, when there are missing ratings, we require that at least one of the group members has rated that item. Hence, when generating a sub-dataset of ratings, by removing ratings with a certain probability, we avoided the cases when an option was not rated by at least one group member. 

\begin{figure}[!htbp]
    \centering
    \includegraphics[width=\linewidth]{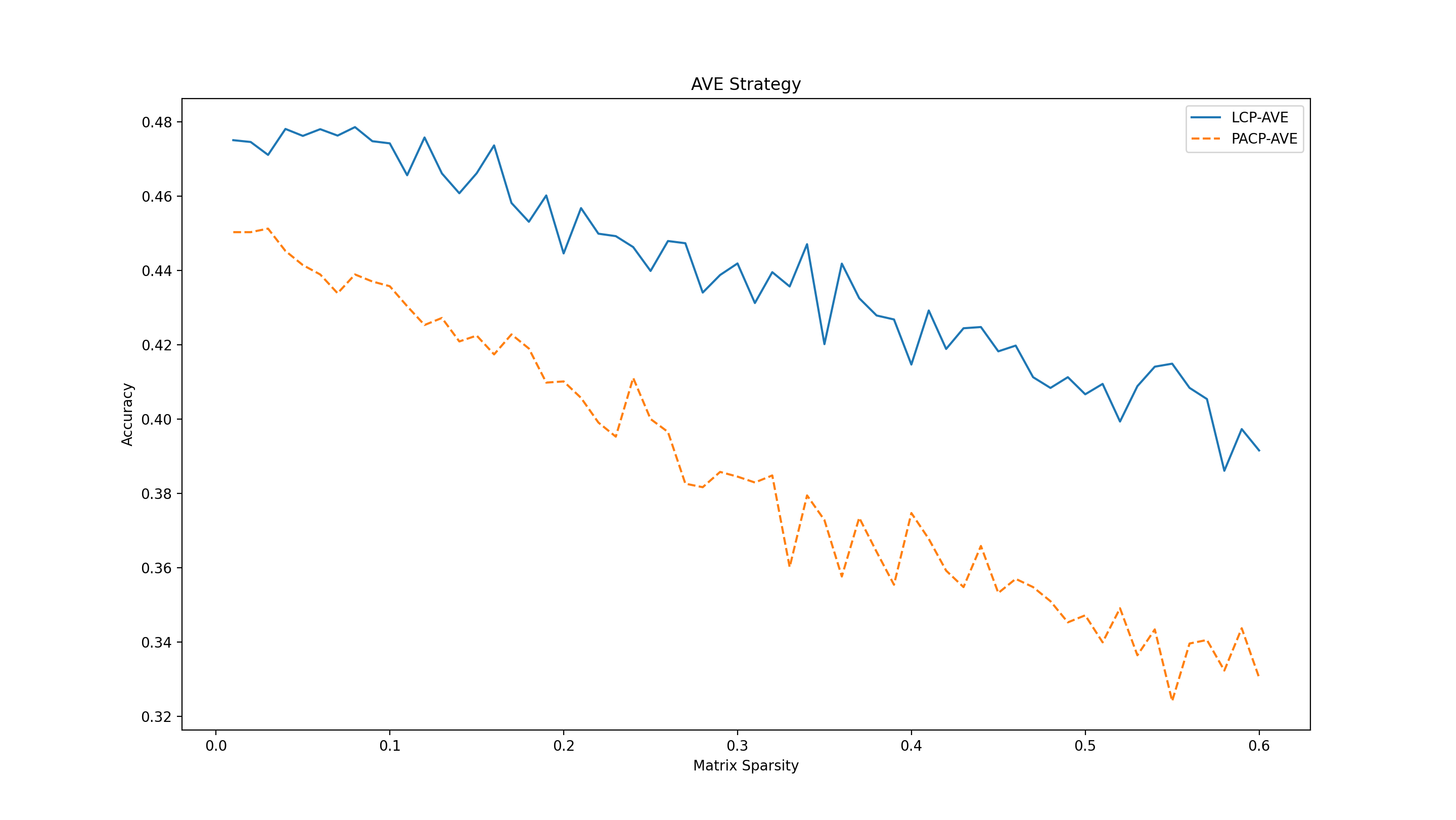}
    \caption{PACP and LCP accuracy in predicting groups' choices when some of the group members' ratings are not available (sparse user-option matrix). The x-axis in this figure indicates the actual sparsity of the generated user-option matrix, i.e., the percentage of ratings that were not present in the rating matrix.}
    \label{fig:sparcity}
\end{figure}

Figure \ref{fig:sparcity} shows the accuracy of PACP-AVE and LCP-AVE for different sub-datasets generated by an increasing probability to remove an existing rating in the original data set. The x-axis in this figure indicates the actual sparsity of the generated user-option matrix, i.e., the percentage of ratings that were not present in the rating matrix that was used to compute the plotted accuracy. To avoid creating randomly a very good or very bad matrix, we repeated this experiment 50 times and reported the average accuracy for each obtained sparsity level. As expected, the accuracy of LCP-AVE and PACP-AVE  decreases as the sparsity of the user-option rating matrix increases. However, it is clear, that LCP-AVE has a better performance than PACP-AVE in dealing with missing data. 
Figure~\ref{fig:sparcity} also shows that with increasing sparsity, the performance gap between LCP-AVE and PACP-AVE grows. It is worth mentioning that we also tested other variants of PACP and LCP, for instance, PACP-MULT, but we do not show these results as they are very similar to those shown here. 

\subsection{Human Ability in Predicting the Group Choice}
\label{Sec:HA}

We have shown that LPC can achieve an accuracy of almost 50\% in predicting the group choice from the observation of the group members' preferences (see Figure~\ref{Square_Standard_Strategies_font}). To better judge the absolute quality of this ML-based prediction, it is interesting to understand whether a human may be better than LCP in predicting the choice of a group, by having the knowledge of the group members' preferences, i.e., their ratings for the alternative options.

\begin{figure}[!htbp]
    \centering
    \includegraphics[width=0.7\linewidth]{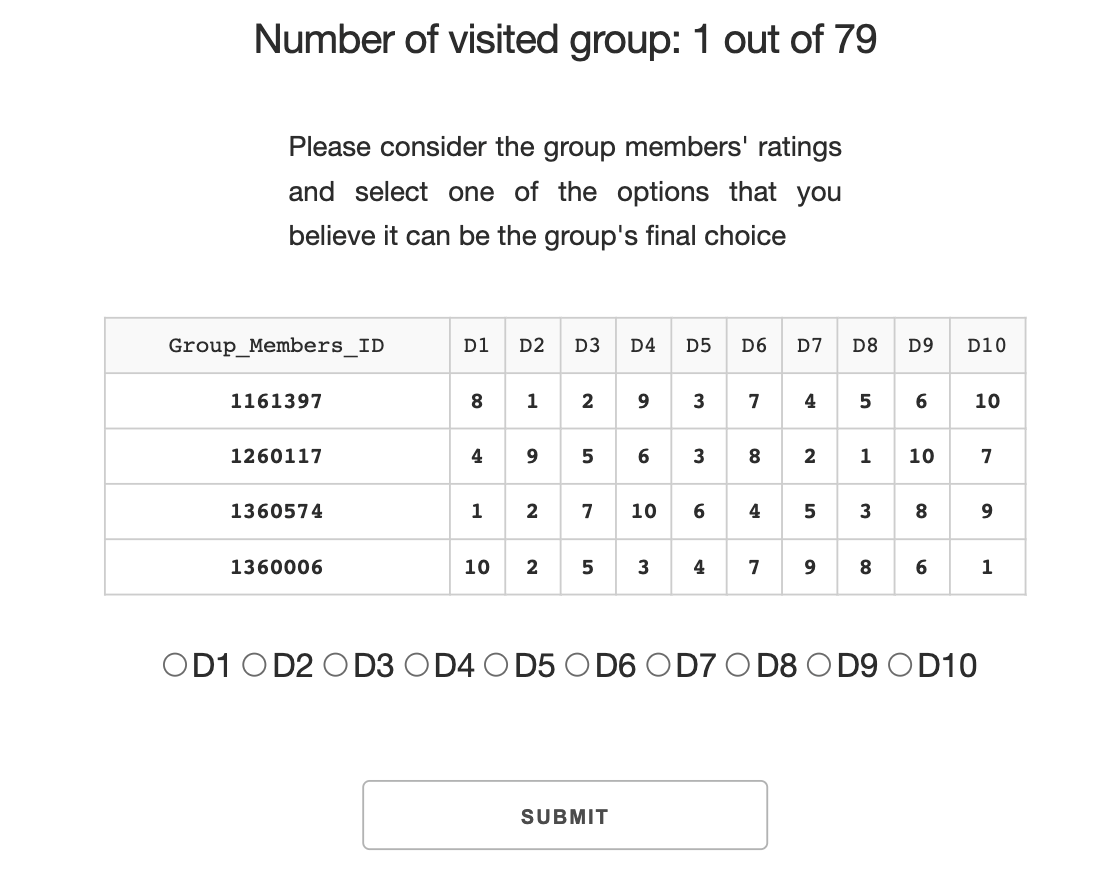}
    \caption{User study GUI where the group members' ratings for ten options are displayed and the subject can select the option that they believe could be the group's final choice.}
    \label{fig:GUI}
\end{figure}

To this end, following a similar approach adopted in~\cite{MASTHOFF04}, we have conducted a user study and asked ten participants to predict the likely group's choice after having observed the group members' ratings. The participants were computer science master students and colleagues at the Free University of Bolzano. We gave the participants a simple task: {\it please consider the group members' ratings shown here and select the option that you believe it can be the group's final choice}. We have implemented a simple GUI (Figure \ref{fig:GUI}), in which the system, for each group in the data set, shows the group members' ratings for ten destinations (D1, ..., D10). Note that no information about the identity of the ten destinations was given and the subjects were not even aware of the fact that it was a destination selection problem faced by the groups. The study participant was requested to select the destination/option that they believed could have been the group's final choice.  

The participants predicted all the group choices of the 79 groups in our dataset. We did not specify any time limitation for the participants and they had to do the whole experiment in one session. All of our ten subjects performed the required task in one session and we did not exclude any of them from our report. On average, they required about 20 minutes to predict all the groups' choices. The average accuracy of the participants in predicting the group choices was $0.37$, with a minimum of $0.28$ and a maximum of 0.46. The variance of the accuracy score was $0.05$.

By comparing these results to the performance of LCP and PACP, shown in Figure~\ref{Square_Standard_Strategies_font}, one can see that human (average) accuracy is much lower than the (average) accuracy obtained by the best LCP variants, and even the accuracy of the best PACP variants (AVE, MULT, COPE). Only the best-performing human (0.46) is approaching the (average) performance of the best LCP and PACP variants, which is obtained with the AVE preference aggregation strategy. This result further confirms the benefit of using our ML predictive models, namely LCP, to solve this task.

\subsection{Using Data Augmentation in LCP}

\begin{figure}[!htbp]
    \centering
    \includegraphics[width=0.9\linewidth]{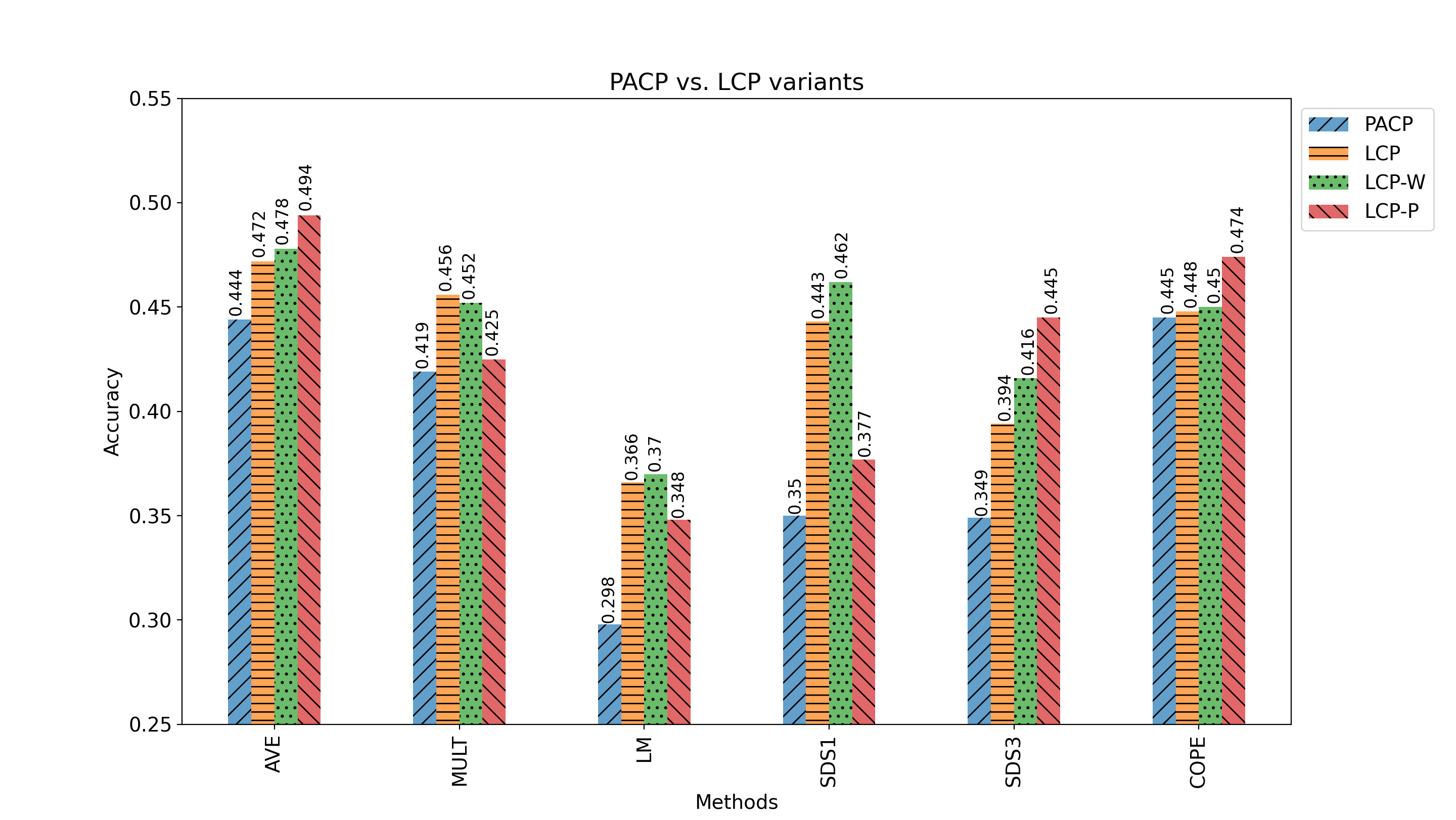}
    \caption{Comparison of the accuracy of LCP and PACP variants with and without data augmentation. As shown in this figure, LCP-AVE-P outperforms the other LCP and PACP variants.}
    \label{Square_Standard_Strategies_font_variant}
\end{figure}

We now consider the LCP variants that use the proposed data augmentation approach: \textit{Winners} and \textit{Permutations} (see Section \ref{sec:data_augm}). Figure \ref{Square_Standard_Strategies_font_variant} shows the performance of the PACP and the LCP variants either with and without synthetic group profiles (\textit{Winners} or \textit{Permutations}). We discuss here our second hypothesis: \textit{in order to cope with the data scarcity of group choices, it is possible to use data augmentation methods, relying on synthetic group choices, and further improve the quality of the proposed group choice prediction method}. 

We observe that the advantage of using the data augmentation methods is not uniform for all the group profile types, i.e., built by using the various preference aggregation strategies that we have considered. Adding \textit{Winners} group profiles to the data set brings a value to all the LCP variants, with the exception of MULT. Interestingly the benefit of the \textit{Winners} data augmentation is more evident for the SDS1 and SDS3 variants. In conclusion, the \textit{Winners} data augmentation seems to be applicable without any fear to deteriorate the LCP performance.

Adding instead \textit{Permutations} data improves the accuracy of LCP-AVE, LCP-SDS3, and LCP-COPE. But \textit{Permutations} are not beneficial at all when the group profiles are built with the other preference aggregation strategies. It is, however, important to note that LCP-AVE-P is the best-performing choice prediction method. Hence, there is a clear value of both data augmentation techniques, but the \textit{Permutations} data augmentation approach requires to be validated before being applied in combination with a specific preference aggregation strategy, case by case.

In order to better understand the effect of adding the synthetic groups in the \textit{Permutations} to the training set, we analyzed the distribution of the predictions over the 10 possible options and compared it to the actual distribution of the group choices in the data set. We found that \textit{Permutations} help to reduce the KL-divergence between the distribution of the predicted choices and the distribution of the actual group choices as it is shown in Table \ref{tab:kl_diverg}. KL-divergence is a statistical metric that measures to what extent two probability distributions are different from each other. The KL-divergence of the LCP variants that use \textit{Permutations}, namely LCP-P, is much lower than that of the original LCP, and the LCP variants that use the data set augmented by the \textit{Winners} group profiles. Moreover, it is also lower than the Kl-divergence of the PACP model. Therefore, these synthetic group profiles aid LCP to generate a choice distribution that is more similar to the observed choice distribution. 

\begin{figure}[!htbp]
    \centering
    \includegraphics[width=1.0\linewidth]{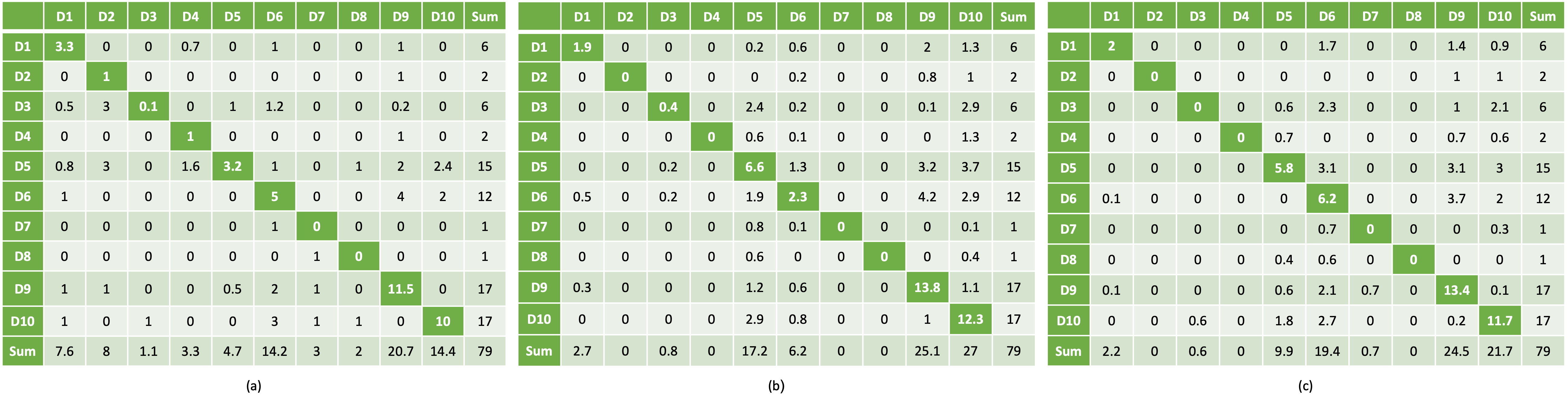}
    \caption{Confusion matrix for PACP-AVE (a), LCP-AVE (b), and LCP-AVE-P (c). The last column indicates the number of actual group choices for each class (destination) and the last row indicates the number of predicted group choices for each class (destination).}
    \label{fig:confusion}
\end{figure}

We further illustrate this finding by comparing the confusion matrices of the PACP-AVE (Figure \ref{fig:confusion} (a)), LCP-AVE (Figure \ref{fig:confusion} (b)), and LCP-AVE-P (Figure \ref{fig:confusion} (c)). In these matrices, each row corresponds to an option chosen by a number of groups indicated in the last column, and the entries in the row show how the predictions for those groups are distributed along the 10 options. By looking at the last (bottom) row of these tables (summary distribution of all the predictions), it is clear that the choice predictions of LCP-AVE concentrate more around the four most popular options ($o_5$, $o_6$, $o_9$, and $o_{10}$) than the PACP-AVE does. While with the addition of \textit{Permutations}, the LCP-AVE-P model produces a (predicted) choice distribution more evenly distributed. Still, by comparing the bottom row of the LCP-AVE and LCP-AVE-P matrices with the last column (the true distribution of the group choices) one can immediately see the bias of the learning methods, which predict more often the popular group choices.

\begin{table*}[t]
\centering
\caption{KL-divergence between the predicted choice distribution and the actual choice distribution. A smaller value indicates a more similar distribution of the predicted choices to the actual choices.
}
\label{tab:cinfusion}
\scalebox{1.0}{
\begin{tabular}{| r | c | c | c | c | }
\hline
 & PACP & LCP & LCP-W & LCP-P\\
\hline
AVE & 0.202 & 0.212 & 0.278 & {\bf 0.196}\\
\hline
MULT & 0.251 & 0.184 & 0.282 & {\bf 0.164}\\
\hline
LM & 0.293 & 0.372 & 0.569  & {\bf 0.212}\\
\hline
\end{tabular}
}
\label{tab:kl_diverg}
\end{table*}

\subsection{Results Significance}
We also tested the significance of all our results with the Wilcoxon sign-rank test \cite{wilcoxon1992individual}. 
Table \ref{tbl:significance} shows the significance level (p-value) of the improvement made by LCP-* compared to PACP-* and LCP-*-P compared to LCP-*. LCP-*-W was never significantly better than LCP.

\begin{table}[!htbp]
\centering
\caption{Significance level (p-value) of the learning method LCP-*-P improvement in comparison to LCP}\label{tbl:significance}
\scalebox{0.8}{
\begin{tabular}{|c|c|}
\hline
\textbf \textbf{Compared Variants} &  \textbf{p-value} \\
\hline
LCP-LM vs. PACP-LM & $0.048$\\
\hline
LCP-SDS1 vs. PACP-SDS1 & $0.049$ \\
\hline
LCP-AVE-P vs. LCP-AVE& $0.019$ \\
\hline
LCP-SDS3-P vs. LCP-SDS3 & $0.011$ \\
\hline
LCP-COPE-P vs. LCP-COPE & $0.020$  \\
\hline
\end{tabular}
}
\end{table}

%% file: 07_Discussion.tex
\section{Discussion and Future Work}
\label{sec:conclusion}

{\bf Group Choice Prediction}

\noindent GRSs have focused on the problem of how to properly aggregate and use individual preferences in order to suggest items that a group will be happy to choose. In this article, by considering that preference aggregation techniques can also be leveraged to predict the choice that a group could or should make, we have hypothesized that the performance of these strategies can be improved by the adoption of a proper machine learning approach (see Hypothesis 1 in Section~\ref{sec:hyp}) and even further improved by using a proper data augmentation method (see Hypothesis 2 in Section~\ref{sec:hyp}). Our hypotheses are motivated by the Social Decision Scheme (SDS) theory that explicitly focuses on the group choice prediction problem. SDS models how the collective response (group choice), which is the result of possibly complex inter-group interactions that happen during the group decision-making process, can be generated by relying only on individual preferences (group members' preferences).

\vspace{1ex}
\noindent {\bf Group Profiles and Machine Learning}

\noindent We have proposed a new approach to group choice prediction that learns a mapping from the group profile to the group choice. The group profile is a summary representation of the individual group members' preferences and can be generated by any preference aggregation strategy. We have shown that our approach, named LCP (Learning-based Choice Prediction), produces effective predictions, significantly more accurate than those generated by the usage of standard aggregation strategies (PACP). 

We have considered a range of aggregation strategies in assessing the quality of LCP: Average, Multiplicative, Least Misery  SDS1, SDS3, and COPE. We have found that group profiles constructed by using the Average preference aggregation strategy enable LCP to produce the best results in our data set. We have also shown that the better performance of LCP, compared to  PACP baseline, which was firstly assessed on a dense rating matrix, is maintained when only partial knowledge about individual preferences of the members of the group is available, i.e., only some of the individual preferences are known. The empirical analysis also indicates that as the sparsity of the rating matrix increases, the performance gap between LCP and PACP  increases as well, with a steeper performance decline for  PACP.  Moreover, in a user study, we have shown that the proposed choice prediction method, LCP, is more accurate than human assessors trying to predict the group choice from the observation of the group members' preferences (ratings for the options).

Our group choice prediction approach uses a set of observed groups, along with the group members' ratings for the considered options and the consequent choice that the group made, in order to train the group choice predictive model. Hence, our approach is dependent on the size and quality of the training data. In order to cope with the data scarcity problem, i.e., having only a limited number of observed groups with their choices, we have proposed two training data augmentation methods. These methods are grounded on assumptions about the properties of the target map from group profiles to group choices. The first assumption is that if all the group members select one item as their preferred option, then this must be the group's final choice. Group profiles with that characteristic were called \textit{Winners}. Then in the second method, by assuming that groups are influenced in their decision by the relative scores of the options in the group profile, and not the actual options, we have defined \textit{Permutations} profiles and added them to the training data. We have found that \textit{Winners} are never damaging the prediction accuracy of LCP, but their benefit is small (as very few group profiles are added by this method). While, in combination with certain types of group profiles (those produced by AVE, SDS3, and COPE preference aggregation strategies) \textit{Permutations} do help LCP to produce more accurate predictions. Moreover, it is observed that this result is obtained by generating a distribution of the predicted group choices more similar to the true (observed in the data) group choice distribution.  

\vspace{1ex}
\noindent {\bf Limitations and Future Work}

\noindent We must acknowledge some limitations of our choice prediction approach and its evaluation. Firstly, in this study, we have used a single data set and of rather small size. We have however shown in our experiments that the performance of the proposed method on our data set, which is actually the combination of two data sets referring to different group decision problems, is comparable to the performance of the same method on a single data set. This shows that the proposed approach can be robust to solve a combination of prediction tasks, provided that the number of options is the same (see Section~\ref{sec:combinig_data}). In fact, the advantage of our approach is that it is not using any specific characteristic of the choice prediction task, apart from the knowledge of the individual preferences. This clearly makes the proposed approach general.

However, we recognize that there is an urgent need to test the application of LCP to other data sets, in order to further increase the generality of our results. Therefore, in the future, it is important to consider other, possibly larger, data sets. The data sets must be generated by collecting group choices in a broad range of application domains (e.g., tourism, music, video). However, it is worth noting that data sets with information about the individual preferences and the corresponding group choice are not available now, and new observational studies, such as that described in~\cite{GROUPDM18}, are needed to produce such data sets. 

Secondly, the proposed choice prediction technique is currently designed to tackle decision problems with a small number of options; in our data set, ten options were available for the group to choose from. It is surely necessary to design and analyze methods that can scale to a larger number of options. This may be achieved with different types of learning algorithms and with more effective modeling approaches to summarize the preferences of the group members. In that respect, it could be important to derive from the surveyed ML-based group recommender systems approaches, alternative solutions to model the joint preferences of a group, as those based on hidden features of the group.

A third limitation is related to the fact that a rating data set, which models individual preferences, could be very sparse, even sparser than the data sets considered in Section~\ref{sec:sparse_ratings}.  So, an appropriate method to deal with these situations is needed. We note that in our current solution of the problem we require that for each option there must be at least one group member that has rated it. An extension of LPC to deal with sparser rating data must be designed. One line of research could be to test the effect of replacing missing ratings with ratings obtained from a rating prediction method, and then to construct group profiles based on both real and predicted ratings.

\vspace{1ex}
\noindent {\bf Conclusion}

\noindent In conclusion, the extensive analysis that we present in this paper has clearly indicated the effectiveness and practical applicability of the proposed methods for group choice prediction.

This information, the likely choice that a group will make, can be immediately used to generate a recommendation, namely, by recommending the predicted choice, hence helping the group to faster converge to a decision. However, by having the knowledge of the likely choice of the group, the recommender system can also leverage it to generate other types of recommendations, for instance, presenting items similar to the predicted choice but with additional important properties, such as items that are more novel or fairer choices. Hence, we believe that addressing the group choice prediction problem can open the research on novel and interesting group recommendation techniques and especially conversational approaches, which can greatly benefit from the prediction of the likely choice of the group, to better interact with the group members in supporting their decision-making process.

Hence, notwithstanding its limitations, we believe that the proposed approach represents a concrete tool for better understanding groups, their discussions, and to generate more compelling GRSs.

%% file: templateArxiv.bbl
\begin{thebibliography}{10}

\bibitem{ricci:2022:ch1}
Francesco Ricci, Lior Rokach, and Bracha Shapira.
\newblock {\em Recommender Systems: Techniques, Applications, and Challenges.}, pages 1--45.
\newblock Springer, Boston, MA, USA, 2022.

\bibitem{MASTHOFF2022}
Judith Masthoff.
\newblock Group recommender systems: Beyond preference aggregation.
\newblock In Francesco Ricci, L.~Rokach, and B.~Shapira, editors, {\em Recommender Systems Handbook}, pages 381--420, New York, NY, USA, 2022. Springer.

\bibitem{JAMESON04}
Anthony Jameson.
\newblock More than the sum of its members: challenges for group recommender systems.
\newblock In {\em Proceedings of the Working Conference on Advanced Visual Interfaces}, AVI ’04, page 48–54, New York, NY, USA, 2004. Association for Computing Machinery.

\bibitem{stasser1999}
Garold Stasser.
\newblock A primer of social decision scheme theory: Models of group influence, competitive model-testing, and prospective modeling.
\newblock {\em Organizational Behavior and Human Decision Processes}, 80(1):3--20, 1999.

\bibitem{MASTHOFF04}
Judith Masthoff.
\newblock Group modeling: Selecting a sequence of television items to suit a group of viewers.
\newblock In {\em Personalized digital television}, pages 93--141. Springer, Dordrecht, Netherlands, 2004.

\bibitem{DELIC2018GDM}
Amra Delic, Julia Neidhardt, and Hannes Werthner.
\newblock Group decision making and group recommendations.
\newblock In {\em 2018 IEEE 20th Conference on Business Informatics (CBI)}, volume~1, pages 79--88, Vienna, Austria, 2018. IEEE.

\bibitem{FORSYTH14}
Donelson~R Forsyth.
\newblock {\em Group dynamics}.
\newblock Cengage Learning, Boston, MA, USA, 2018.

\bibitem{wong2016}
Sebastien~C. Wong, Adam Gatt, Victor Stamatescu, and Mark~D. McDonnell.
\newblock Understanding data augmentation for classification: When to warp?
\newblock In {\em 2016 International Conference on Digital Image Computing: Techniques and Applications (DICTA)}, pages 1--6. IEEE, 2016.

\bibitem{felfernig2018group}
Alexander Felfernig, Ludovico Boratto, Martin Stettinger, and Marko Tkal{\v{c}}i{\v{c}}.
\newblock {\em Group recommender systems: An introduction}.
\newblock Springer Cham, Gewerbestrasse 11, 6330 Cham, Switzerland, 2018.

\bibitem{ardissono2003intrigue}
Liliana Ardissono, Anna Goy, Giovanna Petrone, Marino Segnan, and Pietro Torasso.
\newblock Intrigue: personalized recommendation of tourist attractions for desktop and hand held devices.
\newblock {\em Applied artificial intelligence}, 17(8-9):687--714, 2003.

\bibitem{zhiwen2005adaptive}
Yu~Zhiwen, Zhou Xingshe, and Zhang Daqing.
\newblock An adaptive in-vehicle multimedia recommender for group users.
\newblock In {\em 2005 IEEE 61st Vehicular technology conference}, volume~5, pages 2800--2804. IEEE, 2005.

\bibitem{cook1978priority}
Wade~D Cook and Lawrence~M Seiford.
\newblock Priority ranking and consensus formation.
\newblock {\em Management Science}, 24(16):1721--1732, 1978.

\bibitem{dong2021}
Yucheng Dong, Yao Li, Ying He, and Xia Chen.
\newblock Preference--approval structures in group decision making: Axiomatic distance and aggregation.
\newblock {\em Decision Analysis}, 18(4):273--295, 2021.

\bibitem{cao2018AGREE}
Da~Cao, Xiangnan He, Lianhai Miao, Yahui An, Chao Yang, and Richang Hong.
\newblock Attentive group recommendation.
\newblock In {\em The 41st International ACM SIGIR Conference on Research \& Development in Information Retrieval}, SIGIR '18, page 645–654, New York, NY, USA, 2018. Association for Computing Machinery.

\bibitem{Sankar2020GroupIM}
Aravind Sankar, Yanhong Wu, Yuhang Wu, Wei Zhang, Hao Yang, and Hari Sundaram.
\newblock Groupim: A mutual information maximization framework for neural group recommendation.
\newblock In {\em Proceedings of the 43rd International ACM SIGIR Conference on Research and Development in Information Retrieval}, SIGIR ’20, page 1279–1288, New York, NY, USA, 2020. ACM.

\bibitem{sen1977social}
Amartya Sen.
\newblock Social choice theory: A re-examination.
\newblock {\em Econometrica: journal of the Econometric Society}, 45(1):53--89, 1977.

\bibitem{friedkin_johnsen_2011}
Noah~E. Friedkin and Eugene~C. Johnsen.
\newblock {\em Models of Group Decision-Making}, page 235–258.
\newblock Structural Analysis in the Social Sciences. Cambridge University Press, New Yor, NY 10013-2473, USA, 2011.

\bibitem{lakshmanan2020machine}
Valliappa Lakshmanan, Sara Robinson, and Michael Munn.
\newblock {\em Machine learning design patterns}.
\newblock O'Reilly Media, 1005 Gravenstein Highway North, Sebastopol, CA 95472, 2020.

\bibitem{antoniou2017data}
Antreas Antoniou, Amos Storkey, and Harrison Edwards.
\newblock Data augmentation generative adversarial networks, 2018.

\bibitem{shorten2019survey}
Connor Shorten and Taghi~M Khoshgoftaar.
\newblock A survey on image data augmentation for deep learning.
\newblock {\em Journal of Big Data}, 6(1):60, 2019.

\bibitem{GROUPDM18}
Amra Delic, Julia Neidhardt, Thuy~Ngoc Nguyen, and Francesco Ricci.
\newblock An observational user study for group recommender systems in the tourism domain.
\newblock {\em Information Technology \& Tourism}, 19(1-4):87--116, 2018.

\bibitem{delic2018social}
Amra Delic, Judith Masthoff, Julia Neidhardt, and Hannes Werthner.
\newblock How to use social relationships in group recommenders: empirical evidence.
\newblock In {\em Proceedings of the 26th Conference on User Modeling, Adaptation and Personalization}, pages 121--129, New York, NY, United States, 2018. ACM.

\bibitem{GROUP_DM16}
Amra Delic, Julia Neidhardt, Thuy~Ngoc Nguyen, Francesco Ricci, Laurens Rook, Hannes Werthner, and Markus Zanker.
\newblock Observing group decision making processes.
\newblock In {\em Proceedings of the tenth ACM conference on Recommender systems, RecSys'16}, page 147–150, New York, NY, USA, 2016. Association for Computing Machinery.

\bibitem{Yiannakis1992}
Andrew Yiannakis and Heather Gibson.
\newblock Roles tourists play.
\newblock {\em Annals of tourism Research}, 19(2):287--303, 1992.

\bibitem{GIBSON02}
Heather Gibson and Andrew Yiannakis.
\newblock Tourist roles: Needs and the lifecourse.
\newblock {\em Annals of tourism research}, 29(2):358--383, 2002.

\bibitem{NEIDHARDT2014}
Julia Neidhardt, Reiner Schuster, Leonhard Seyfang, and Hannes Werthner.
\newblock Eliciting the users' unknown preferences.
\newblock In {\em Proceedings of the 8th ACM Conference on Recommender systems}, pages 309--312, 2645767, 2014. ACM.

\bibitem{gretzel2004}
Ulrike Gretzel, Nicole Mitsche, Yeong-Hyeon Hwang, and Daniel~R Fesenmaier.
\newblock Tell me who you are and i will tell you where to go: Use of travel personalities in destination recommendation systems.
\newblock {\em Information Technology \& Tourism}, 7(1):3--12, 2004.

\bibitem{moscardo1996}
Gianna Moscardo, Alastair~M Morrison, Philip~L Pearce, Cheng-Te Lang, and Joseph~T O'Leary.
\newblock Understanding vacation destination choice through travel motivation and activities.
\newblock {\em Journal of Vacation Marketing}, 2(2):109--122, 1996.

\bibitem{masthoff2006pursuit}
Judith Masthoff and Albert Gatt.
\newblock In pursuit of satisfaction and the prevention of embarrassment: affective state in group recommender systems.
\newblock {\em User Modeling and User-Adapted Interaction}, 16(3):281--319, 2006.

\bibitem{venables2013modern}
William~N Venables and Brian~D Ripley.
\newblock {\em Modern applied statistics with S-PLUS}.
\newblock Springer Science \& Business Media, Berlin, Germany, 2013.

\bibitem{ripley2007pattern}
Brian~D Ripley.
\newblock {\em Pattern recognition and neural networks}.
\newblock Cambridge University Press, Cambridge, UK, 2007.

\bibitem{chang2011libsvm}
Chih-Chung Chang and Chih-Jen Lin.
\newblock Libsvm: A library for support vector machines.
\newblock {\em ACM transactions on intelligent systems and technology (TIST)}, 2(3):1--27, 2011.

\bibitem{wilcoxon1992individual}
Frank Wilcoxon.
\newblock Individual comparisons by ranking methods.
\newblock In {\em Breakthroughs in Statistics: Methodology and Distribution}, pages 196--202. Springer, New York, USA, 1992.

\end{thebibliography}
